\documentclass[twocolumn]{aastex631}

\usepackage[whole]{bxcjkjatype} 

\usepackage{color}
\usepackage{amsmath, amssymb}
\usepackage[normalem]{ulem} 
\usepackage{multirow}
\usepackage{hyperref}
\newcommand{\co}[2]{CO $J=$ #1 -- #2}
\newcommand{\sio}[2]{SiO $J=$ #1 -- #2}

\newcommand{\rin}{{r_\mathrm{in}}}
\newcommand{\rout}{{r_\mathrm{out}}}

\newcommand{\bb}{{\mathbf{b}}}
\newcommand{\vv}{{\mathbf{v}}}

\newcommand{\jv}{{\mathbf{j}}}

\newcommand{\Cfunc}{{\mathcal{C}}}
\newcommand{\Lfunc}{{\mathcal{L}}}
\newcommand{\omv}{{\text{\boldmath$\omega$\unboldmath}}}
\newcommand{\fv}{{\mathbf{f}}}
\newcommand{\vhat}[1]{{\text{\boldmath$\hat{#1}$\unboldmath}}}
\newcommand{\lfrac}[2]{{{#1}/{#2}}}

\newcommand{\cm}{{\mathrm{\,cm}}}

\newcommand{\second}{{\mathrm{\,s}}}

\newcommand{\km}{{\mathrm{\,km}}}

\newcommand{\kms}{{\km\second^{-1}}}

\newcommand{\au}{{\mathrm{\,au}}}

\newcommand{\pc}{{\mathrm{\,pc}}}

\newcommand{\yr}{{\mathrm{\,yr}}}

\newcommand{\gm}{{\mathrm{\,g}}}

\newcommand{\gram}{{\gm}}

\newcommand{\msunNSP}{{M_\sun}}
\newcommand{\msun}{{\,\msunNSP}}
\newcommand{\lsun}{{L_\sun}}

\newcommand{\solarmass}{\msun}
\newcommand{\solarmassyr}{{\solarmass\yr^{-1}}}

\newcommand{\K}{{\mathrm{\,K}}}

\newcommand{\tirho}{{\tilde{\rho}}}
\newcommand{\tib}{{\tilde{b}}}
\newcommand{\MA}{{M_\mathrm{A}}}
\newcommand{\alphab}{{\alpha_b}}
\newcommand{\ib}{_i}

\defcitealias{shu1991}{SRLL}
\defcitealias{shang2020}{Paper I}
\defcitealias{shang_PII}{Paper II}
\defcitealias{shang_PIII}{Paper III}
\defcitealias{shang_PIV}{Paper IV}
\newcommand{\Synline}{{\textbf{\textsl{Synline}}}}

\accepted{December 27, 2022}
\shorttitle{A Unified Model for Bipolar Outflows from Young Stars: Apparent Magnetic Acceleration}
\shortauthors{Shang et al.}

\begin{document} 
\title{A Unified Model for Bipolar Outflows from Young Stars:\\ Apparent Magnetic Jet Acceleration}

\author[0000-0001-8385-9838]{Hsien Shang （尚賢）}
\affiliation{Institute of Astronomy and Astrophysics, Academia Sinica,  Taipei 10617, Taiwan}

\author[0000-0001-5557-5387]{Ruben Krasnopolsky}
\affiliation{Institute of Astronomy and Astrophysics, Academia Sinica,  Taipei 10617, Taiwan}

\author[0000-0002-1624-6545]{Chun-Fan Liu（劉君帆）}
\affiliation{Institute of Astronomy and Astrophysics, Academia Sinica, Taipei 10617, Taiwan}

\correspondingauthor{Hsien Shang}
\email{shang@asiaa.sinica.edu.tw}


\begin{abstract}

We explore a new, efficient mechanism that can power toroidally magnetized jets up to two to three times their original terminal velocity after they enter a self-similar phase of magnetic acceleration. Underneath the elongated outflow lobe formed with a magnetized bubble, a wide-angle free wind, through the interplay with its ambient toroid, is compressed, and accelerated around its axial jet. The extremely magnetic bubble can inflate over its original size, depending on the initial Alfv\'en Mach number $\MA$ of the launched flow. The shape-independent slope $\partial{}v_r/\partial{}r=2/3t$ comes as a salient feature of the self-similarity in the acceleration phase. Peculiar kinematic signatures are observable in the position--velocity (PV) diagrams and can combine with other morphological signatures as probes for the density-collimated jets arising in the toroidally-dominated magnetized winds. The apparent second acceleration is powered by the decrease of the toroidal magnetic field but operates far beyond the scales of the primary magnetocentrifugal launch region and the free asymptotic terminal state. Rich implications may connect the jets arising from the youngest protostellar outflows such as HH 211 and HH 212 and similar systems with parsec-scale jets across the mass and evolutionary spectra.

\end{abstract}

\keywords{\href{http://astrothesaurus.org/uat/1966}{Magnetohydrodynamical simulations (1966)}; \href{http://astrothesaurus.org/uat/1857}{Astronomical simulations (1857)}; \href{http://astrothesaurus.org/uat/1964}{Magnetohydrodynamics (1964)}; \href{http://astrothesaurus.org/uat/1569}{Star formation (1569)}; \href{http://astrothesaurus.org/uat/1635}{Stellar wind bubbles (1635)}; \href{http://astrothesaurus.org/uat/1636}{Stellar winds (1636)}; \href{http://astrothesaurus.org/uat/870}{Jets (870)}; \href{http://astrothesaurus.org/uat/1390}{Relativistic jets (1390)}; \href{http://astrothesaurus.org/uat/1347}{Radio jets (1347)}; \href{http://astrothesaurus.org/uat/1607}{Stellar jets (1607)}; \href{http://astrothesaurus.org/uat/1576}{Stellar-interstellar interactions (1576)}}

\section{Introduction}
\label{sec:intro}

Astrophysical jets are ubiquitous in accretion systems, from young stars to supermassive black holes \citep[e.g.,][]{ray2021,romero2021}. They are energetic and highly collimated and remain so over large physical scales and dynamic ranges. While the detailed physics of their processes may change from one to another, jets display similarities in their general characteristic features. Although the dynamical mechanisms for their large-scale acceleration and collimation remain enigmatic and unsettled, the generality of the phenomena and fundamental physical processes from the birth to death of stars across the mass spectrum is recognized.

Jets in young stars are known to be signposts of the stars' birth in deeply embedded stellar nurseries \citep[e.g.,][]{ray2021,frank2014}. They persist from the earliest Class 0 to Class II phases in molecular to atomic lines \citep[e.g.,][]{pascucci2022_PPVII}. Jets appear as mass-loss phenomena connected deeply with mass accretion in the protostellar disk \citep[e.g.,][]{lee2020}. Their high velocity and highly collimated morphology indicate origins deep in the gravitational potential well of the surrounding accretion disk \citep[e.g.,][]{shu2000,konigl2000,frank2014,bally2016}. They arise in various forms in magnetocentrifugal winds driven by the Blandford--Payne (BP) mechanism from the inner disk regions \citep[e.g.,][]{blandford1982,shu2000,konigl2000,ray2021}.
This mechanism produces a signature large-scale cylindrically stratified profile of the density and toroidal field component (see below in Section \ref{sec:method}). The X-wind model \citep{shu1994,shu1994b} generates the most salient signatures \citep{shu1995,shang1998}, from the innermost edge of the protostellar disk, by interacting with the stellar magnetosphere \citep{shu1994,ostriker1995}.

Standard theories on the formation and propagation of astrophysical jets consider the initial launch and evolution of freely propagating flows in accretion--ejection systems \citep[e.g.,][]{mckee2007,pudritz2012,ray2021}.
In the context of protostellar jets, the initial launch refers to the magnetocentrifugal BP mechanism from the original radial location in the disk, followed by passing through the critical points of sonic, Alfv\'enic, and fast-magnetosonic transitions before reaching the asymptotic state and collimation of streamlines \citep[e.g.,][]{konigl2000,shu2000}. The flow velocities reach the terminal values determined by the conversion of the rotation and magnetic energy obtained from accretion into the kinetic energy of the gas \citep[e.g.,][]{blandford1982,shu1994,spruit1996_proc}. Beyond the fast-magnetosonic transition, in the asymptotic regime, \citet{shu1995} demonstrate that the initial gradual acceleration continues to reach its ultimate terminal velocity and flow collimation logarithmically provided the physical scale is large enough.

Protostellar jets are intimately connected to large-scale outflows, the presence of which is the outcome of the underlying driver interacting with the ambient environment \citep[e.g.,][]{bally2016,bachiller1996,lee2020}. 
Highly collimated outflows surrounding narrow molecular jets with extremely high velocity (EHV) are commonly associated with deeply embedded Class 0 protostars. The influence and feedback of the ambient environment on the overall system need to be included as an integral part of the theoretical treatment, especially for the earliest phase of star formation.
\citet[][hereafter \citetalias{shang2020} and \citetalias{shang_PII}]{shang2020,shang_PII} advance the physics of such interplay for a jet-bearing hydromagnetic wind interacting with its ambient envelope.
They conclude that outflows formed as wind-driven bubbles should be ubiquitous and are an inevitable integrative outcome of the interplay.
Within the magnetized bubbles, the compressed and shocked regions can be thick and extended, and nested multicavities can form naturally as part of the process. Recent high-angular-resolution and -sensitivity observations of nested-shell structures in several young outflows such as HH 212, DG Tau B, and HH 30 indeed support the formation mechanism of interconnected multicavities.

Within the integrative platform, a new secondary regime of magnetic acceleration of the flow can be identified, which appears to be a prominent feature arising from the magnetic interplay within confined shocked zones. The unique acceleration mechanism explored in this work differs from the primary one that launches and accelerates from the base region of the free magnetocentrifugal wind. We emphasize this specialized form of acceleration is most salient inside the context of very strongly magnetized self-similar elongated bubbles that take the functional form adopted.

We focus on this new phenomenon in this Letter because of its importance and generality in the context of our current framework. 
We highlight the ubiquity of energetic jets from astrophysical systems, their formation theory, and where our work will improve the theoretical picture. In Section \ref{sec:method}, we review the theoretical framework advanced in \citetalias{shang2020} and the formulation established for this series of works.
We summarize the definitions of relevant variables and derive characteristic properties of the acceleration. In Section \ref{sec:numerical}, we demonstrate the numerical examples and their consistency with the analytic expectations. We discuss and summarize the significance and implications of the findings in Sections \ref{sec:discussions} and \ref{sec:summary}.

\section{Theoretical Considerations}
\label{sec:method}

We advance an integrated framework of outflows unifying the jet, the wide-angle wind, and a magnetized nonspherical (spindle-shaped) bubble elongated along the jet axis in \citetalias{shang2020}. Contrary to the expectations of a thin-shell model with self-similar expansion \citep[e.g.,][]{shu1991,KM1992b} and latitude-dependent velocity variations, the bubbles formed in this integrated framework possess extended internal fine structures that arise due to the magnetic interplay of the magnetized wind and the ambient medium.

We demonstrated in \citetalias{shang2020} the physics of the presence of complex structures when a hydromagnetic wind interacts with an ambient toroid. These outflows are the interaction between an ambient medium and an advanced state of an asymptotic wind. 
The hydromagnetic wind can be simplified as a free constant-velocity radial flow with a strong density stratification of $1/\varpi^2$. Through the interaction with the ambient medium, nested structures form surrounding the free wind, first enclosed by a reverse shock as an inner cavity, followed by an extended region filled with compressed wind, then covered by a layer of compressed ambient medium and compressed ambient poloidal field. Between the compressed wind and the ambient medium lies the tangential discontinuity, which is subject to substantial shear and unstable to Kelvin--Helmholtz instabilities (KHI)\@. When the wind is toroidally magnetized, the compressed and shocked regions can become very extended and nonuniform due to the nonlinear growth of the KHI, further enhanced by vorticity generated by magnetic forces. The magnetic forces feed back into the flow velocity, resulting in magnetic pseudopulses in the extended compressed-wind region and giving an impression of bow shocks along the jet axis. 

In the current framework, we find time-dependent structures arising in a compressed-wind region confined by shocks, which are self-similar with respect to the general features of outflow morphology and main kinematics. The elongated bubble establishes its self-similarity prior to the onset of the second acceleration through decay in the pressure of a toroidal magnetic field. This is different from a hypothetical magnetic tower mechanism, which would have to operate from the base of the launching process near the disk.
In this work, we focus on the time-dependent self-similar conditions of the confined wind, which amplify the acceleration given by the decay of the magnetic energy term. We also study the increase of outflow size in interaction conditions out of a steady state for dynamical expansion.

\begin{figure*}
\plotone{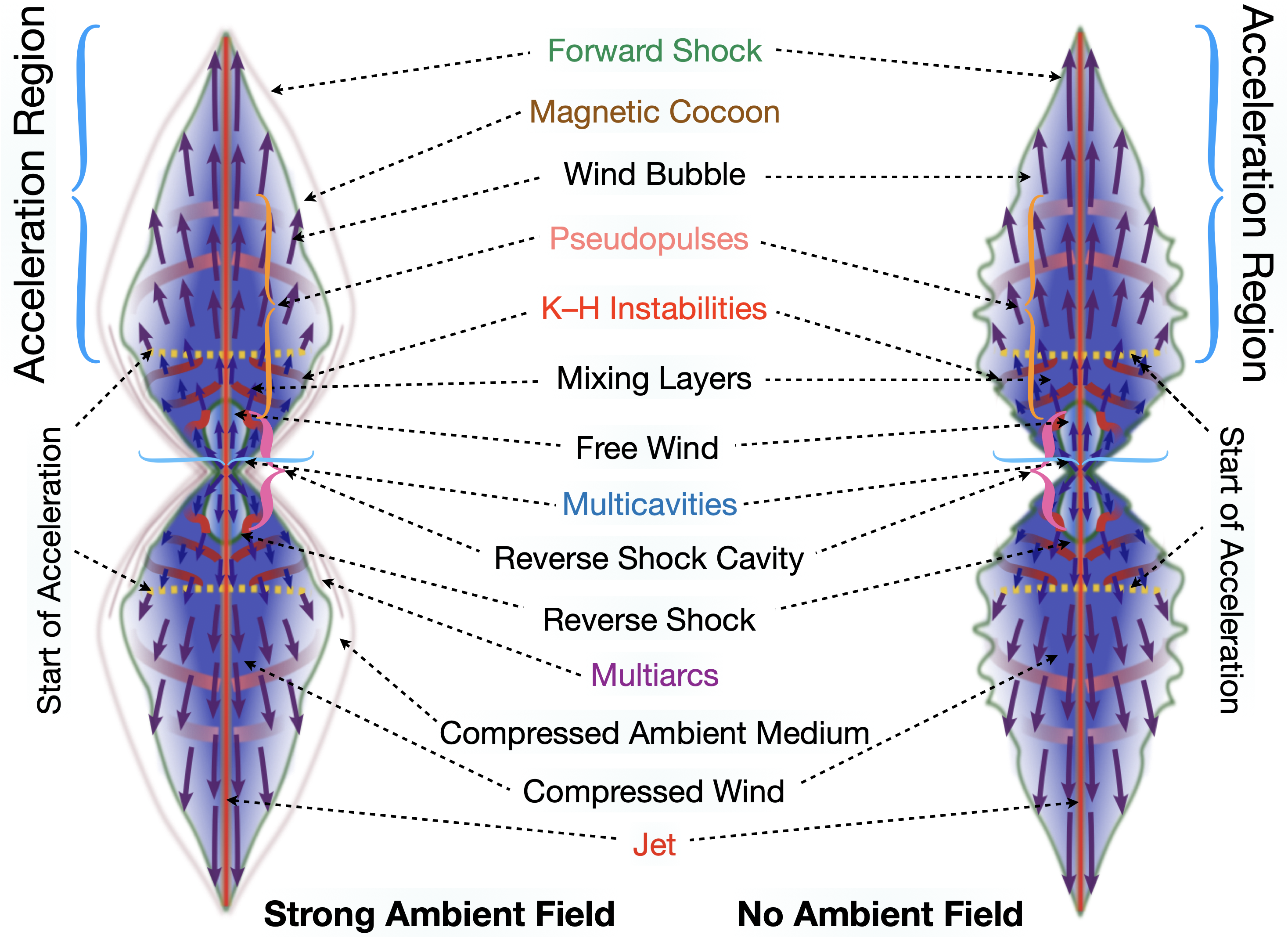}
\caption{Schematic plots of the outflow lobes with magnetic acceleration for a case with a strong ambient magnetic field (left) and without an ambient magnetic field (right).}\label{fig:1}
\end{figure*}

This phenomenon of secondary acceleration occurs beyond the region of compressed wind with complex fine structures of KHI and magnetic pseudopulses, but before the terminal forward shock, within the zone bounded by the reverse and forward shocks. This zone of new physics, where the jet accelerates further beyond the terminal velocity, overlaps with the ``tip'' region in \citetalias{shang2020}. 
In Figure \ref{fig:1}, we illustrate our schematic configuration for the second acceleration. In this figure, we focus on the regime of physics discussed specifically in this work, different from those considered in Papers I and II, for the magnetically dominated bubbles ($1< \MA \lesssim 6$) and their structures. We will discuss where this newly observed magnetic acceleration occurs in the configurations and their implications.

The foundation and the associated computational and analytical methodologies utilized are shown in \citetalias{shang2020}, and their kinematic features in \citetalias{shang_PII}. For consistency, we follow the same methodology as previous works in the following subsections.

\subsection{The Magnetized Wind}
\label{subsec:windMHD}

We consider a magnetocentrifugally driven wind, dominated by the wound-up toroidal component well beyond the transition region to reach its final asymptotic state, such as one considered in \citet{shu1995} (for a comparison, see the end of this subsection). The toroidal velocity of the wind declines steadily at large distances due to the conservation of angular momentum. The free wind reaches approximately a poloidal flow with a purely toroidal field frozen inside, $\bb_p=0$ and $v_\phi=0$, with mass density $\rho=\rho_\mathrm{wind}$ and gas pressure $p=a_{\mathrm{wind}}^2 \rho$. Within the spherical coordinates in axisymmetry, the velocity ($v_r$, $v_\theta$, $v_\phi$) and magnetic field ($b_r$, $b_\theta$, $b_\phi$), a simplified set of MHD equations was derived in \citetalias{shang2020}. A summary of variables and symbols is given in Table \ref{tab:quantities}.
 
In a sufficiently cold wind, steady-state equations can be written as Equation (24)--(27) in Section 2.2 of \citetalias{shang2020}, which
admit solutions that give these three constants $v_r$, $\rho\varpi^2$, and $|b_\phi|\varpi$ for $p=0$. These properties give the cylindrically stratified density profile $\rho\propto\varpi^{-2}$, an exactly radial flow of velocity $\vv=v_r\vhat{r}$, and magnetic field $\bb=b_\phi\vhat{\phi}$, for the freely propagating wind before it encounters the ambient medium.

For an axisymmetric $b_\phi$ magnetic component, a current function $\Cfunc=-\varpi b_\phi$, proportional to the total current carried in one hemisphere, can be defined and helps in computing the poloidal current density $\jv_p$ and magnetic force $\fv_c$ (Table \ref{tab:quantities}). The force term $\Cfunc^2/2$ behaves similarly to a pressure $\fv_c=-(\rho\varpi^2)^{-1}\nabla\left(\lfrac{\Cfunc^2}{2}\right)$.
In a cold free wind, the value of $\Cfunc$ stays exactly constant, and relevant gradients vanish.

In this work, for mathematical convenience, we define
$\tirho\equiv\rho\varpi^2$, $\tib\equiv b_\phi\varpi=-\Cfunc$, and their ratio
\begin{equation}
\label{eqn:mu}
\mu\equiv{}b_\phi/(\varpi\rho)=\tib/\tirho\ ,
\end{equation}
independent of $\varpi$ for a free wind. This ratio $\mu$ symbolically defines a relative ratio of the magnetic field to mass density. 

The steady-state free wind takes the form of constants of the problem:
\begin{equation}
  \label{eqn:initcon}
  \tirho=D_0, \quad
  \tib=b_0, \quad
  v_{r}=v_{0}, \quad
  v_\theta=0\ ,
\end{equation}
where $D_0$, $b_0$, and $v_0$ are constants of density, magnetic field, and velocity, respectively, defined at the inner radial boundary of the problem. Their numerical values are given in Table \ref{tab:quantities}.

The Alfv\'en speed $v_\mathrm{A}=b_\phi/\rho^{1/2}=\tib/\tirho^{1/2}$ is constant in the free-wind region and equal to its inner radial BC value $v_\mathrm{A0}\equiv b_0/D_0^{1/2}$.
In other regions of the flow, downstream from the free wind, the value of $v_\mathrm{A}$ can and does differ from the BC and free-wind value $v_\mathrm{A0}$.
The free wind has an initial Alfv\'en Mach number $\MA=v_0/v_\mathrm{A0}$, for the wind velocity $v_0$ and the initial Alfv\'en speed in the wind $v_\mathrm{A0}$:
\begin{equation}
\label{eqn:wind:MA}
    \MA
    =v_0\sqrt{D_{0}}/{b_0}\ .
\end{equation}
Additionally, 
\begin{equation}
    \Cfunc_0=b_0=\sqrt{D_0}v_0/\MA\,
\end{equation}
and
\begin{equation}
    \label{eqn:acc:e15}
    \mu_0=\left(\frac{\tib}{\tirho}\right)_\mathrm{BC}=\frac{b_0}{D_0}=\frac{v_0}{\MA\sqrt{D_0}}\ ,
\end{equation}
are set exactly in the boundary condition. These values of $\MA$, $v_0$, wind density $\rho_\mathrm{w}$, and $D_0$ fix the scale and value of the toroidal magnetic field in the wind. Within the cold free wind, $\Cfunc=\Cfunc_0$ and
$\mu=\mu_0$.

This overall character applies to strongly magnetized wind launched from a sufficiently close vicinity to the innermost portion of the disk, irrespective of the origin of its magnetic field, provided that the proper jet density and $b_\phi$ can maintain their profiles for the features constructed and remain sufficiently radial $v_r \gg v_\theta$. The launch region also needs to be much smaller than the physical scales within which the very elongated bubble has reached its self-similar phase (\citetalias{shang2020, shang_PII}).

The initial free-wind flow in this work is assumed to have reached the superfast asymptotic regime in \textit{steady state} before interacting with an ambient medium and becoming enclosed by the reverse shock. 
The free asymptotic steady-state X-wind studied by \citet{shu1995} continues the magnetocentrifugal acceleration process beyond its fast point, considering carefully the inner $\theta$ boundary conditions and the expected magnetic behavior, including the slow logarithmic decay of magnetic energy toward more considerable distances.
In its native format, \citet{shu1995} provide an asymptotic description in which magnetocentrifugal effects
proceed together with magnetic torque along each wind streamline
(summarized in Equation (29) of \citetalias{shang2020} through $C^X\beta^X$, where $C^X$ decays as $1/\log(r)$, despite $C^X$ sharing a similar role to the $\Cfunc$-function adopted here).
When such a wind is used as a BC to study outflows, it is equivalent to cutting off the magnetocentrifugal process at a specific scale corresponding to a particular value of $C^X$\@. The Alfv\'en Mach number $\MA^X$ is defined equivalently in Equation (30) of \citetalias{shang2020}, showing the effect of $C^X\beta^X$, the magnetic Maxwell torque per unit of mass flux. The regime of $1<\MA\lesssim 6$ in this work is magnetically dominated, as explained there, and the $C^X\beta^X$ term is relatively large, allowing a smaller $\MA$. The wind magnetization labeled as $\MA$ in this series reflects a parameterized set of wind configurations in space and time.
The resulting magnetically dominated self-similar bubble driven by a highly magnetized wind is out of steady state and dynamically expanding for our purposes.

\subsection{Equations of Self-Similarity}
\label{subsec:eqns_selfsim}

We write down a simplified form of equations for the free- and compressed-wind regions, ignoring terms due to gas pressure, $b_p$, $v_\theta$, and $v_\phi$. Under those approximations, the equations of time evolution for $v_r$, $\rho$, and $b_\phi$ simplify to
\begin{equation}
    \label{eqn:acc:e1}
    \partial_t v_r+v_r\partial_r v_r+ \frac{\Cfunc\partial_r\Cfunc}{\rho\varpi^2}=0\ ,
\end{equation}
\begin{equation}
    \label{eqn:acc:e3}
    \partial_t(r^2\rho)+\partial_r(r^2 v_r\rho)=0\ ,
\end{equation}
\begin{equation}
    \label{eqn:acc:e4}
    \partial_t(r b_\phi)+\partial_r(r v_r b_\phi)=0\ .
\end{equation}

Equations for density and magnetic field can be written (using $\tirho$, $\tib$, and $D_t\equiv\partial_t+\vv\cdot\nabla=\partial_t+v_r\partial_r$) as
\begin{equation}
    \label{eqn:acc:e3tilde}
    \partial_t\tirho+\partial_r(v_r \tirho)=D_t\tirho+\tirho\partial_r v_r=0\ ,
\end{equation}
\begin{equation}
    \label{eqn:acc:e4tilde}
    \partial_t\tib+\partial_r(v_r \tib)=D_t\tib+\tib\partial_r v_r=0\ .
\end{equation}
Experimentally observed in our numerical results reported previously (\citetalias{shang2020, shang_PII}), self-similarity applies to the majority of the flow, either free or compressed, except for the exact KHI features at any instant of time. 

Here we note the property that for any quantity $A=A(x)$ behaving (to sufficient approximation) self-similarly as a function of $x\equiv r/v_0 t$, its total derivative in a region of \textbf{self-similar acceleration} can be estimated as
\begin{equation}
\label{eqn:acc:e20}
D_t A=\partial_t A+v_r\partial_r A\approx(-r/t+v_r)\partial_r A
\equiv v_E\partial_r A   
\end{equation}
while in steady state $D_t=v_r\partial_r$.
Equation (\ref{eqn:acc:e20}) also defines the velocity $v_E\equiv v_r-r/t$, pervasive in time-dependent self-similar flows needing to consider total derivatives.

Both steady-state and self-similar formulas agree in a steady-state region such that $r\ll v_r t$ and thus $v_r\approx v_E$. It is the case for the wind in the (usually small) innermost cavity enclosed by the reverse shock.
The observation of the tip region shown in \citetalias{shang2020} as discussed in Appendix \ref{sec:tips} is now connected to this full set of self-similar equations.

In Appendix \ref{sec:mu} we examine these self-similar equations and obtain an important property, Equation (\ref{eqn:acc:e26}), linking the magnetic quantity $\mu\tib=\tib^2/\tirho=b_\phi^2/\rho=v_\mathrm{A}^2$ to the hydrodynamical results. That equation can be fulfilled in an uninteresting manner, $\partial_r v_r=0$, as in the free wind, or more interestingly as
\begin{equation}
\label{eqn:acc:e27}
\mu\tib=(v_r-\frac{r}{t})^2\ ,
\end{equation}
or $v_\mathrm{A}^2=v_E^2$
as most probably can happen in an acceleration region.
Equation (\ref{eqn:acc:e27}) gives an algebraic functional form for the behavior of $b_\phi$ in the acceleration region. It is, however, not to be utilized in a free-wind region.

\subsection{The Acceleration}
\label{subsec:acceleration}

The whole process of acceleration is under the control of the radial force, Equation (\ref{eqn:acc:e1}). The acceleration originates in the magnetic force term which converts toroidal magnetic field to radial velocity and can be rewritten as
\begin{equation}
\label{eqn:force_r}
\frac{\tib\partial_r\tib}{\tirho}=\mu\partial_r\tib=\partial_r(\mu\tib)-\tib\partial_r\mu\ ,
\end{equation}
The accelerating effects of this force are amplified by the $\partial_t v_r$ term of Equation (\ref{eqn:acc:e1}).

The magnetic force term
in the acceleration region can be simplified to $\mu\partial_r \tib\approx\partial_r(\mu\tib)$, the first term on the right-hand side of Equation (\ref{eqn:force_r}), by noticing that in that region, $\mu$ is nearly constant, with $D_t\mu\approx v_E\partial_r\mu\approx0$.
This motivates the definition of a function behaving in the acceleration region as a \textbf{specific enthalpy} function for the magnetic force,
\begin{equation}
    \label{eqn:acc:enthapy_func}
    h=\mu\tib\,
\end{equation}
which helps define a Bernoulli constant in Equation (\ref{eqn:acc:f9}) below.
The existence of a function such that $-\nabla h=\jv\times\bb/\rho$ is special. When present, the generation of vorticity by magnetic fields $\nabla\times(\jv\times\bb/\rho)$ becomes zero (curl of a gradient), consistent with the acceleration region but not with the pseudopulses \citepalias{shang_PII}.

It may be illustrative to compare this magnetic specific enthalpy function with that of an ideal gas. 
Here we have a specific enthalpy $\mu\tib=b_\phi^2/\rho=v_\mathrm{A}^2$ equal to twice the magnetic pressure per unit mass $b_\phi^2/2\rho$, and equal to the square of the Alfv\'en speed. This relation between specific enthalpy, pressure, and wave speed could compare symbolically with that present in an ideal gas with pressure $p$, sound speed $(\gamma p/\rho)^{1/2}$, and specific enthalpy $(p/\rho)[\gamma/(\gamma-1)]$ with $\gamma=2$.
This analogy (short of a complete equivalence) works for the radial force component $-\partial_r h$ and its relation with gas pressure and wave (Alfv\'en) speed. In Appendix \ref{sec:HD_analog} we have extended it into a simplified model for the radial force, able to reproduce some of the results regarding radial velocity and position behavior of the acceleration region.

The momentum equation
    $D_t v_r + \partial_r{h} =0$,
simplified using $h=v_E^2$ to
\begin{equation}
    v_E\partial_r (v_r + 2v_r-2r/t)=0
\end{equation}
can be fulfilled if either $v_E=0$ (only at the end of the acceleration region after $h$ has been spent), or the gradient of velocity takes the value

\begin{equation}
    \label{eqn:acc:e30}
    \partial_r v_r=\frac{2}{3t}\ ,
\end{equation}
independent of $\MA$. In self-similar variables, the acceleration, 
\begin{equation}
    \label{eqn:acc:e31}
    \frac{\partial(v_r/v_0)}{\partial(r/v_0t)}=\frac{2}{3}\ ,
\end{equation}
is a constant. 

This functional form for the gradient of $v_r$, $\partial v_r/\partial r$, could be ``experimentally'' checked against the strongly magnetized wind regimes for the phenomenological context below.

\subsection{Bernoulli Constant and the Maximum Velocity}
\label{subsec:maxvel}

Inside the acceleration region, a Bernoulli function can be defined by combining three terms
\begin{equation}
    \label{eqn:acc:f9}
    H=\frac{v_r^2}{2}+h-\frac{r^2}{3t^2}\ ,
\end{equation}
where the last term is obtained by using Equation (\ref{eqn:acc:e30}) to integrate $\partial_t v_r\approx-(r/t)\partial_r v_r$.
This quantity is therefore a Bernoulli constant ($\partial_r H=0$) within the acceleration region, beginning at $r\ib$ and ending at $r_f$.
At the beginning of the acceleration region,
\begin{equation}
    \label{eqn:acc:f10}
    H\ib=\frac{v_0^2}{2}+h\ib-\frac{r\ib^2}{3t^2}\ ,
\end{equation}
and at the end
\begin{equation}
    \label{eqn:acc:f11}
    H_f=\frac{v_f^2}{2}-\frac{r_f^2}{3t^2}\ ,
\end{equation}
because the magnetic strength has been spent to zero at the end of the acceleration region.

The initial specific magnetic enthalpy $h\ib$, is calculated at $r\ib$, where
$h\ib=\mu\ib\tib\ib$,
at the intersection extrapolated down through the $2/3$ slope line down to the $v_r/v_0=1$ line (shown in Figure \ref{fig:2} below in Section \ref{subsec:accel_numeric}; for a detailed derivation, see Appendix \ref{sec:accel_zone}.). 

As a simple convenient approximation, however, we consider 
$h\ib\approx h_0=\mu_0 b_0=v_0^2/\MA^2$, leading to the following simple formula:
\begin{equation}
    \label{eqn:acc:f26}
    v_f=v_0(1+2/\MA)\ .
\end{equation}

Equation (\ref{eqn:acc:f17}) implies that $v_f\leq3v_0$; in combination with Equation (\ref{eqn:acc:f26}), this result requires $\MA\geq1$. This is expected because sub-Alfv\'enic flows require a different treatment.
The accelerated self-similar $v_f$ and $r_f$ found in Equation (\ref{eqn:acc:f26}) can be substantially larger than the hydrodynamic momentum-conserving values, now analytically quantified. We demonstrate this analytic result in our numerical computations.

With regard to the validity of the formulation as presented, the trans-Alfv\'enic $\MA=1$ limit is simply symbolic but not realistic because the basic assumption of the wind having reached a superfast asymptotic state has been included in the boundary condition. A full study including wind launching will be implemented into this framework in future work.

\section{Numerical Demonstrations}
\label{sec:numerical}

Simulations setups are described in detail in Papers I and II\@. We summarize key ingredients here for the computational data utilized in this work.

\subsection{Numerical Results}
\label{subsec:results}

The parameter space contains wind cases for seven values of $\MA=1.2$, $1.5$, $2$, $3$, $6$, $18$, and $30$ for $v_0=100\kms$. For each of the free winds, there are four values of $n=1$, $2$, $4$, and $6$, and two levels of ambient magnetization $\alphab=1$ and $0$, at $6400\times2016$ resolution.
Here we check the validity of our derivations of the formulation for the strongly magnetized winds of $\MA=1.2$ to $\MA=30$, most relevant to the regime considered in this work. The cases $\MA=1.2$, $1.5$, $2$, $3$, and $\MA=18$ are newly performed runs using otherwise the same setup and parameter space as those of $\MA=6$ and $\MA=30$ reported in Papers I and II\@.

We note that self-similarity is established and maintained in a wind-blown bubble as long as the ambient medium [Equation (\ref{eqn:rhoabar})] and the steady wind [Equation (\ref{eqn:rhowbar})] follow the same $1/r^2$ pattern, as explained, e.g.,\ in Appendix A of \citet{KM1992b} \citep[see also][]{shu1991}. This result is basically independent of the $\theta$ profile as the $r$ and $\theta$ directions are independently separable. In this tip region, the background tapering $n=0$ toroid (isothermal sphere) profile dominates over the $n>0$ configurations as $n$ becomes bigger.

The exploration of numerical results in this section shows that the acceleration effect is rather invariant and follows the theoretical results of Section \ref{sec:method}.

\begin{figure*}
\centering
\plotone{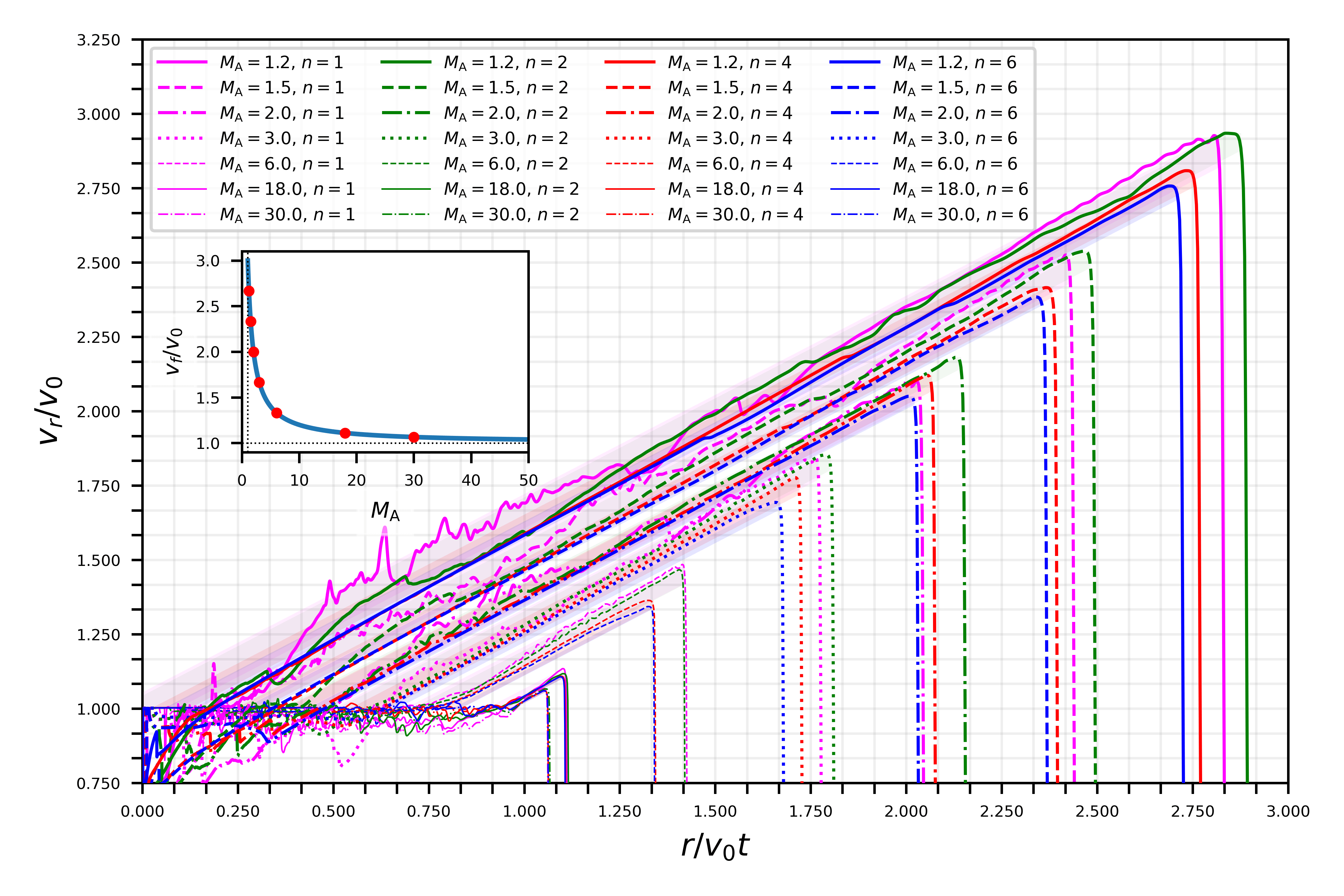}\\
\plotone{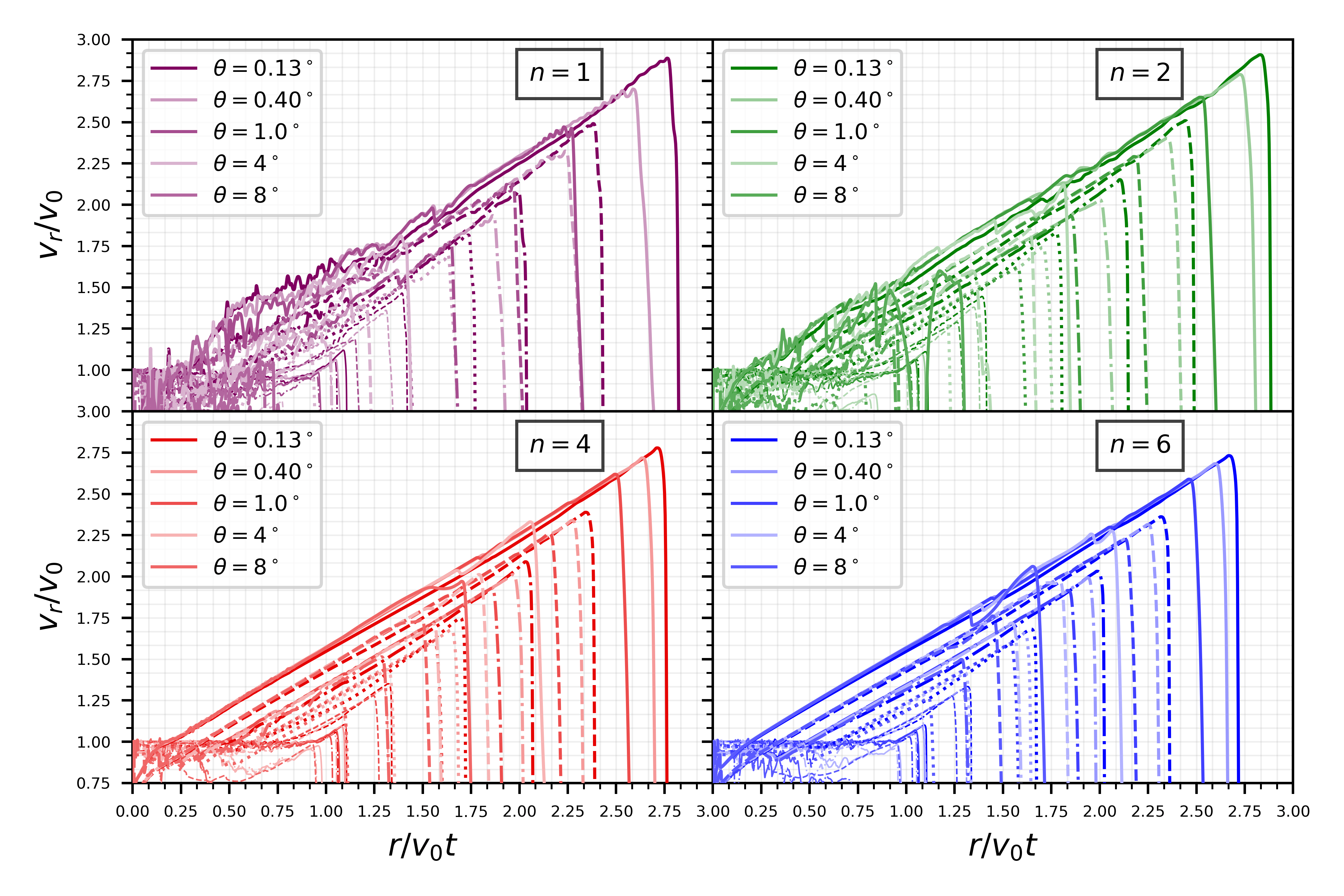}
\caption{{\it Top panel:} radial profile of self-similar acceleration close to the jet axis at $\theta=0\fdg0446$ is shown in self-similar coordinates $v_r/v_0$ vs. $r/v_0t$ for the cases of $\MA=1.2$, $1.5$, $2$, $3$, $6$, $18$, and $30$ winds (distinguished by line styles) with $n=1$, $2$, $4$, and $6$ toroids (distinguished by color hues). The inset shows the plot of $v_f/v_0$ as a function of $\MA$ [Equation (\ref{eqn:acc:f26})]. The $v_f/v_0$ values corresponding to the illustrated $\MA$ values are shown as red dots in the inset.
{\it Lower panel:} radial velocity $v_r/v_0$ vs. $r/v_0t$ profiles for different $\theta$ are shown for the cases with $n=1$, $2$, $4$, and $6$ toroids. 
Data are shown together, for each $n$, for these $\theta$: 0\fdg13, 0\fdg4, 1\arcdeg, 4\arcdeg, and 8\arcdeg, with line styles showing $\MA=1.2$, $1.5$, $2$, $3$, $6$, $18$, and $30$ winds, identical to the top panel.}
\label{fig:2}
\end{figure*}

\subsection{The Velocity Gradient}
\label{subsec:accel_numeric}

We derive the velocity gradient in the self-similar acceleration zone in Equation (\ref{eqn:acc:e31}). We first check if the functional forms for the gradient $\partial v_r/\partial r$ match the numerical results. The top panel of Figure \ref{fig:2} shows the collection of cases from all $n$ values at $\theta=0\fdg0446$ near the jet axis for the maximum radial velocity $v_r$. The constant velocity gradient is observed to be followed by all the cases at this same $\theta$ angle. 
Each of the $v_r$ curves starts from a horizontal portion full of oscillations from the magnetic pseudopulses, followed by the rise of $v_r/v_0$ with a constant slope of $2/3$ to a maximum value before ending just before the terminal forward shock. 
For cases with wider openings in $n=4$ and $6$, the smoother acceleration regions are less affected by magnetic pulses introduced by the interplay. For the stronger magnetization with smaller $\MA$ values, the acceleration often starts below the $v_r/v_0=1$ level directly from a rebound from a noticeable, deep pulse.

Along the $\MA$ values shown, the second apparent acceleration has contributed to the increase of velocity from around the 10\% level for the intermediate $\MA$ range, $18$---$30$, about $1.75$--$2$ for $\MA=2$--$3$, and can shoot up to an approximate factor of $2$--$3$ close to $\MA=1$--$1.2$.
All the cases are below the symbolic limit of $v_r/v_0=3$, implied by Equation (\ref{eqn:acc:f26}), when $\MA=1$. This approximate trend of $v_f/v_0$ is shown in the top panel inset of Figure \ref{fig:2}.

To inspect how the constant slope holds in the 2D acceleration zone,
we sample the radial profiles of all the $n$ and $\MA$ cases of interest for the $\theta$ angles. Self-similar acceleration is observed and reproduced in the 2D space, not limited to the jet axis,
in the bottom panel of Figure \ref{fig:2}.
The peak velocities reached for different $\MA$ values, and $\theta$ values are naturally mingled on each of the panels labeled by $n$, and they all appear to follow the same slope and same trend. The fluctuations in $v_r/v_0$ due to the magnetic pseudopulses are noisier as expected due to the smaller openings in $n=1$ and, to some extent, in $n=2$. They are virtually nonexistent in $n=4$ and $n=6$ in the acceleration zone. 
This indicates that the self-similar condition applied in the derivation is robust in the whole 2D domain of acceleration.

\subsection{Very Magnetic Bubbles}
\label{subsec:magbubbles}

\begin{figure*}
\centering
\epsscale{1.1}
\plottwo{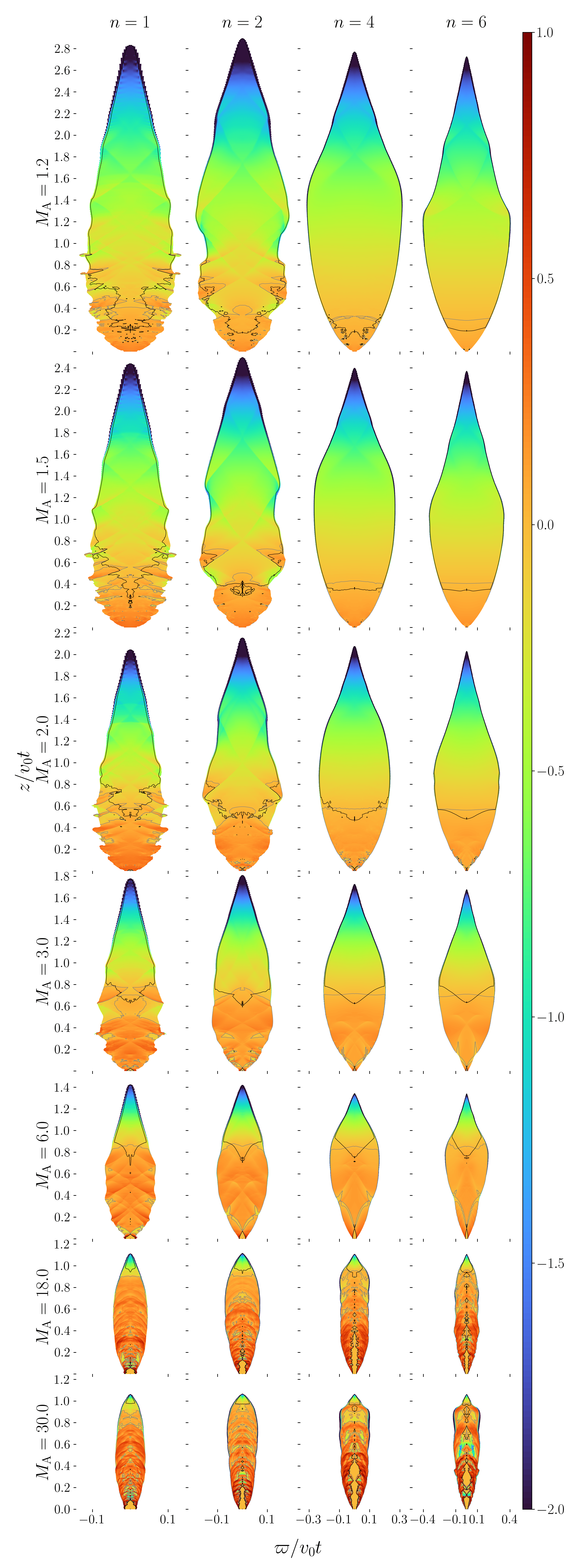}{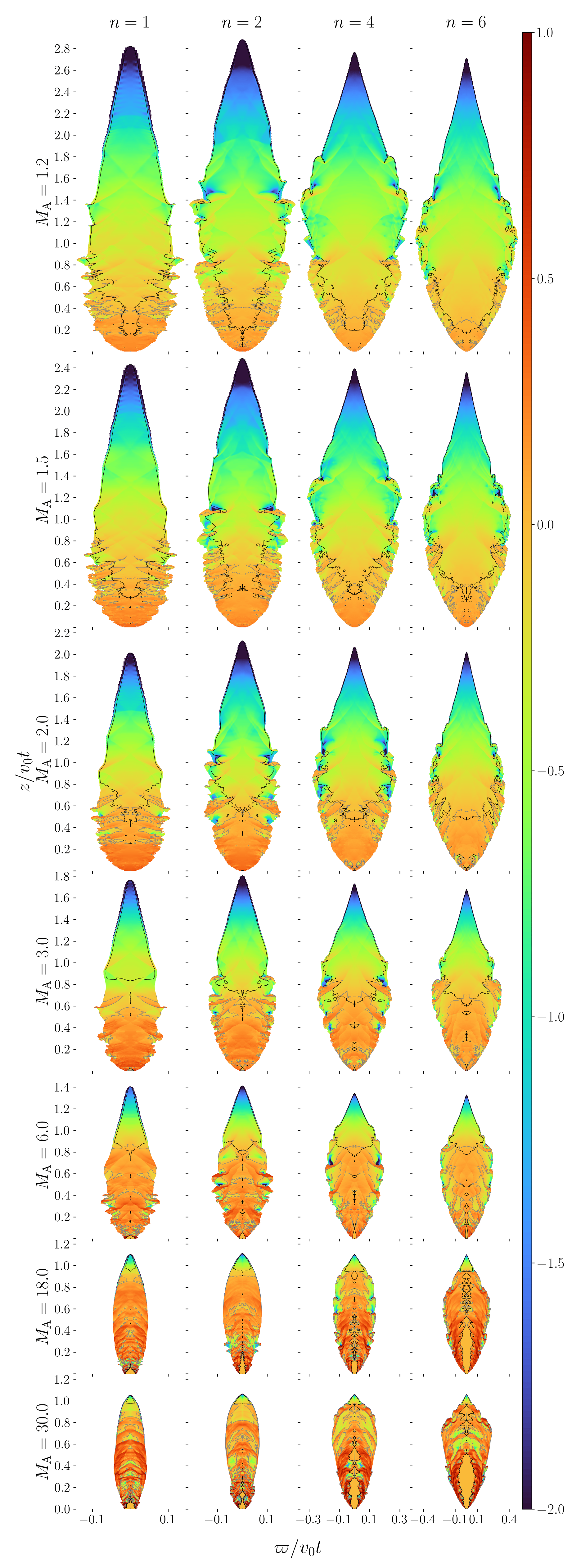}
\epsscale{1.0}
\caption{The 2D spatial profiles of $\log_{10}\left({-\Cfunc}/b_0\right)$ for winds of $\MA=1.2$, $1.5$, $2$, $3$, $6$, $18$, and $30$ with $n=1$, $2$, $4$, and $6$ toroids and ambient poloidal field strengths of $\alphab=1$ (left) and $\alphab=0$ (right), in which $b_0\equiv v_0\sqrt{D_0}/\MA$. The black contours are loci of $v_r/v_0=1$, and the gray contours are loci of $-\Cfunc/b_0=1$. The spatial axes are labeled in units of $v_0t$. The horizontal $\varpi$ axes have been exaggerated by factors of $3.5$, $2.3$, $1.4$, and $1.0$ for $n=1$, $2$, $4$, and $6$, respectively.}
    \label{fig:3}
\end{figure*}

For the primary expected behavior of $b_\phi$ dropping to zero near the tip of the acceleration zone, we naturally look for the variable $\tib$ and normalize to the value $b_0$ at the boundary.
Figure \ref{fig:3} summarizes the situation using the local $\Cfunc$ normalized by $b_0$. The variables
${-\Cfunc}/b_0={\tib/b_0}$
should be $1$ at the inner radial boundary and remain so throughout the uncompressed free-wind region. This quantity traces the strength of $|b_\phi|$ to very small values. The locations where the transitions into the acceleration zone take place are marked by loci of $v_r/v_0=1$.

\begin{figure*}
    \centering
    \plotone{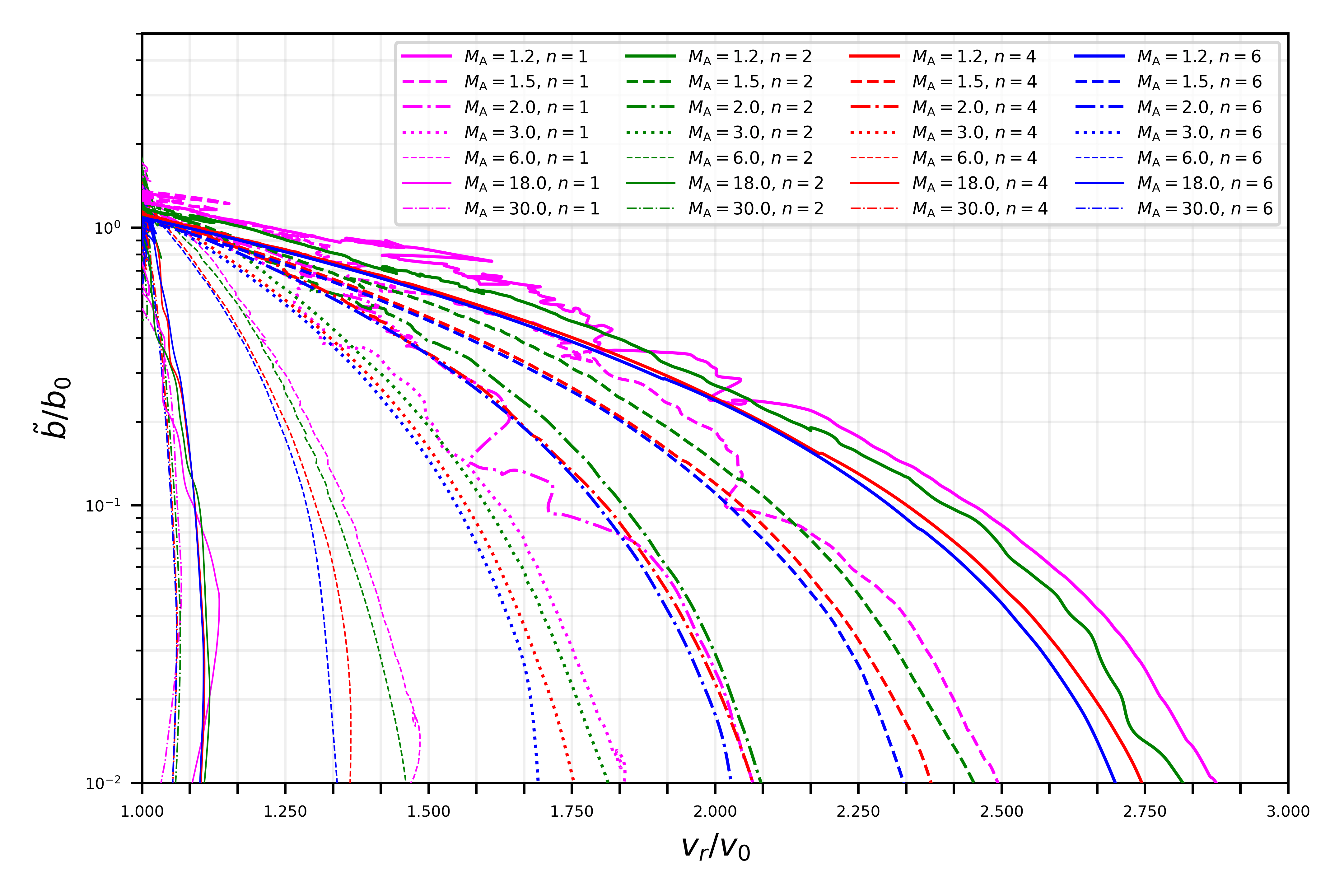}\\
    \plotone{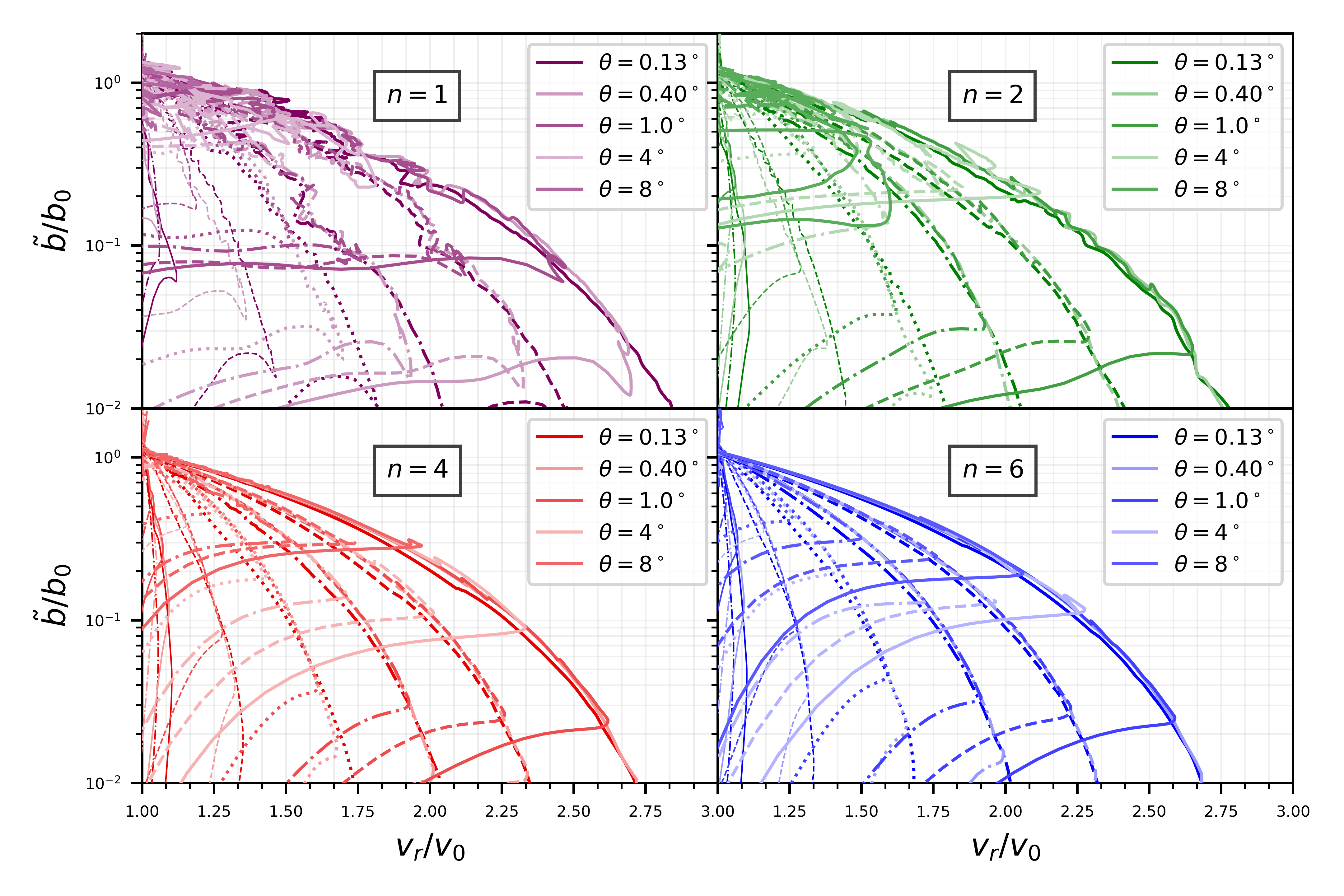}
    \caption{Top: variation along $\theta=0\fdg0446$ of $\tib/b_0=-\Cfunc/b_0$ against $v_r/v_0$. Bottom: $\tib/b_0$ vs. $v_r/v_0$ for $\theta$ of 0\fdg13, 0\fdg4, 1\arcdeg, 4\arcdeg, and 8\arcdeg. The same line styles and colors as in Figure \ref{fig:2} are used.}
    \label{fig:4}
\end{figure*}

We observe how $\tib$ decreases toward the tip of the acceleration region with the value of $\MA$. The acceleration region becomes significantly larger in the extremely magnetically dominated bubbles represented by small $\MA$ values, approaching up to $z/(v_0t)\approx3$ in the axial direction. They demonstrate the characters as magnetic bubbles driven by very strong magnetic pressure. How the decay of $\tib$ powers the acceleration at the far end of the compressed-wind region is directly shown in Figure \ref{fig:4} through the variation of $\tib$ versus $v_r/v_0$.

\begin{figure*}
\centering
\epsscale{1.1}
\plottwo{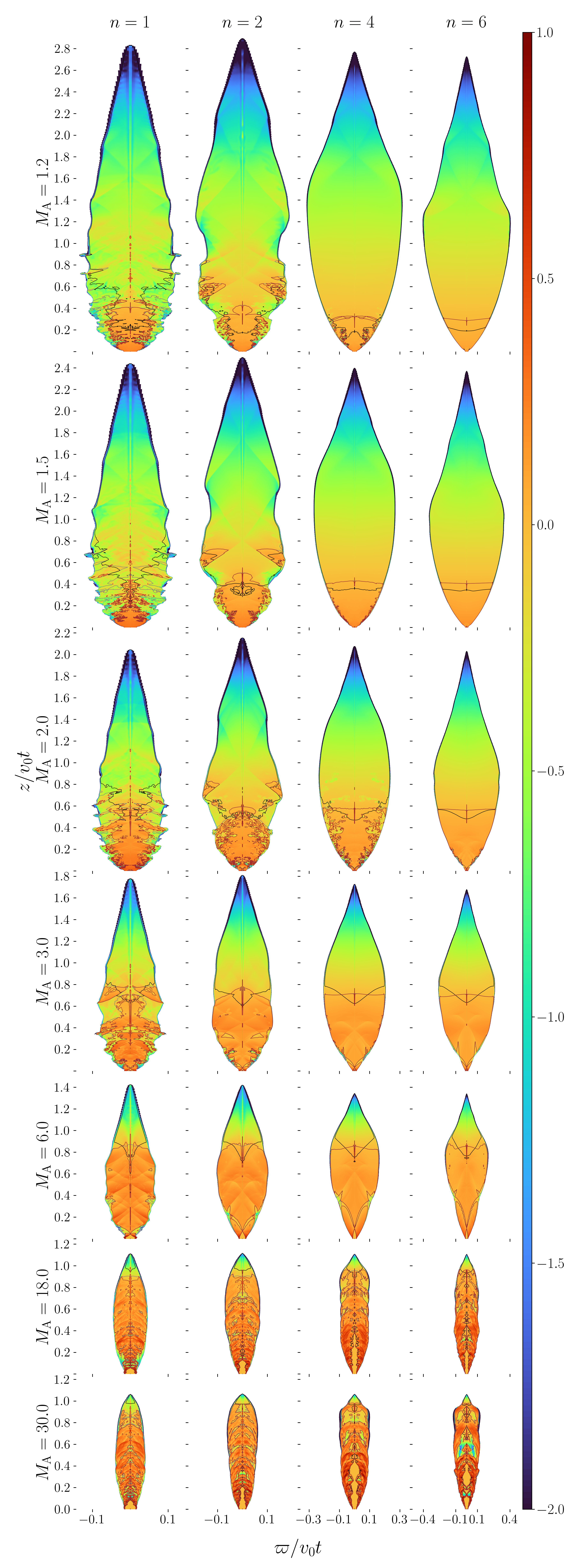}{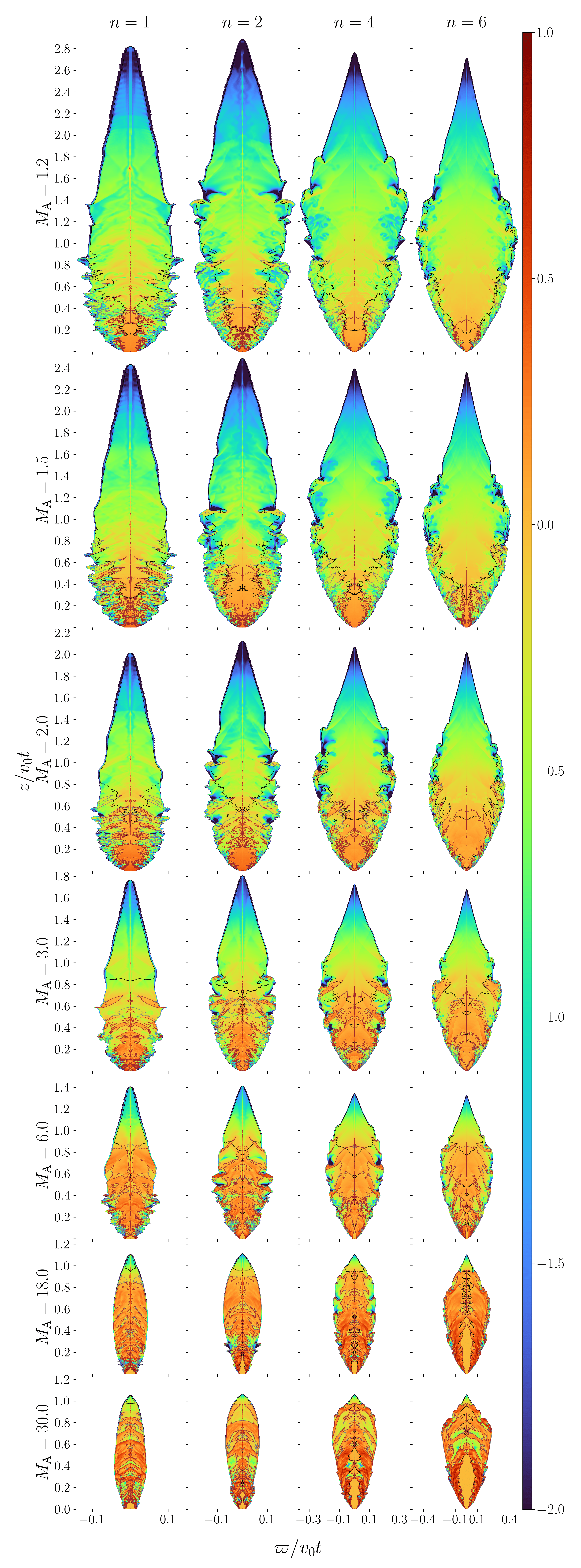}
\epsscale{1.0}
\caption{The 2D spatial profiles of $\log_{10}\left(h/h_0\right)$, a specific magnetic enthalpy within the acceleration region, for winds of $\MA=1.2$, $1.5$, $2$, $3$, $6$, $18$, and $30$ with $n=1$, $2$, $4$, and $6$ toroids and ambient poloidal field strengths of $\alphab = 1$ (left) and $\alphab = 0$ (right), in which $h_0 \equiv v_0^2/\MA^2$. The black contours are loci of $v_r/v_0=1$ and the brown contours are loci of $h/h_0=1$. The spatial axes are labeled in units of $v_0t$, with horizontal $\varpi$ axes exaggerated by factors as in Figure \ref{fig:3}.
\label{fig:5}
}
\end{figure*}

The newly defined specific magnetic enthalpy in Equation (\ref{eqn:acc:enthapy_func}) is shown in Figure \ref{fig:5}, in the form of $h/h_0$,
where $h_0=v_0^2/\MA^2$.
The profiles of $\mu$ are shown in Figure \ref{fig:7} and Appendix \ref{sec:mu}, and they indeed vary very little in the smooth regions. Through the Bernoulli function $H$ and $\partial_r H=0$, the acceleration starts from $v_r/v_0=1$, which often has a higher specific enthalpy than $h/h_0=1$ for most $\MA\lesssim6$.
The function $h=\mu\tib$ is always an important parameter of this flow (equal to the square of the local Alfv\'en speed). However, in this work, it is a specific magnetic enthalpy function only in the acceleration region and with the condition that $\mu\approx$ constant. Regions of magnetic vorticity generation cannot have a magnetic acceleration term derivable from an enthalpy function.

The color contours of $\tib$ trace the magnetic pseudopulses in the compressed wind from right across the reverse shock all the way up to the uppermost gray contour of $\tib/b_0=1$. The magnetic pulses form bow-shaped compressed $\tib$ and give the impression of aligned bow shocks along the jet axis, most evident in $\MA\sim18$--$30$ especially for the smaller $n=1$ and $n=2$ range. Also worth noting is how $\tib$ traces the KHI at the interface of the compressed wind and compressed ambient media on the sides.

The contrast of the KHI patterns at the interfaces of interplay suggests how the local conditions support the growth of different modes. However, these patterns all vanish similarly toward the very top portions of the bubbles. The KHI modes are suppressed in the upper acceleration zone despite their different local growth environments. This is consistent with the inability to generate vorticity because the magnetic force can be expressed as the gradient of a quantity [Equation (\ref{eqn:acc:enthapy_func})] in the acceleration region, similar to the condition arising in the free-wind zone.

The regions where $\mu$ is constant cannot generate magnetic vorticity. The regions where $\mu$ varies help to trace the presence of KHI regions with magnetic vorticity generation, enabling pseudopulses. After that region stops, it is possible to enter the acceleration region, with $h=\mu\tib$ now behaving as a specific enthalpy function. The decay of $\tib$ along the acceleration region, as shown in Figure \ref{fig:3}, powers the acceleration. A detailed comparison of the $\alphab=1$ and $\alphab=0$ panels of these figures shows more prominent pseudopulses for $\alphab=0$, especially for larger $\varpi$ or smaller $n$, due to the presence of KHI\@. This has the effect of delaying acceleration. Horizontal or slightly V-shaped contours of $v_r/v_0=1$ for $\alphab=1$ are often replaced for $\alphab=0$ by sharply V-shaped contours for which a larger $z$ value is needed to enter a smooth acceleration.

One salient feature from the extremely magnetized bubble regime ($\MA\lesssim 3$) is the significantly expanded compressed-wind region overtaking almost the entire lobe volume. The whole bubble structure is powered by a wind that is launched from the disk by, e.g.,\ an X-wind or an inner disk wind, a process taking place in the smaller scales. The large acceleration region (like any other part of the compressed-wind region) is powered by the launching region, but it is not part of it, distinct in this regard from a tower model. This is true even for those extremely magnetized (barely superfast) winds, for which most of the lobe volume is occupied by the apparent second acceleration region transferring magnetic to kinetic energy.

\section{Discussions}
\label{sec:discussions}

We discuss the theoretical and observational implications of our new findings presented in this Letter. We also discuss its generality, limitation, and applicability to other systems.

\subsection{Signature of the Apparent Acceleration}
\label{subsec:accel_presence}

The apparent second acceleration is a prominent signature of the bubble driven by the strongly toroidally magnetized wind under the self-similar elongated bubble environment. It makes a clear kinematic feature, especially on a position--velocity diagram or through proper motions.

In the kinematic signatures of the elongated magnetized bubbles explored in \citetalias{shang_PII}, there indeed appears a new velocity feature in the strongly magnetized systems of $\MA=6$ in which the jet velocity centroids can further shoot up in position--velocity diagrams made of column densities. This is the acceleration signature that motivates the study of this work. PV systems with a larger $\MA$ display smaller tips with fluctuating velocity centroids without shooting up.

This strong jet acceleration appears prominently from the top portion of the compressed-wind region as a linear increase of the velocity centroid if proper excitation of the local density is present. This is a strong effect in contrast to the mild fluctuations around the original ejection velocity caused by the magnetic pulses. In Figure \ref{fig:6} (a) of the PV diagrams of column density, the butterfly-shaped distribution of the column densities stretches out owing to acceleration in both jet portions compared to those of larger $\MA$ values near the bottom panels. The velocity peaks bend out toward the larger absolute velocity, and the wiggles of the magnetic pseudopulses fade moving up the $z$-axis. The bending of the velocity centroids outward starts from lower heights as $\MA$ goes lower. The column densities in the acceleration zone selected with mixed material are significantly reduced as shown in Figure \ref{fig:6} (b), indicating that this zone is powered with the primary jet/wind material.

This acceleration has an exact slope of $2/3$ in our presented figure panels, which are at $45\arcdeg$. We note it is a welcome coincidence for our selected inclination angle. The viewing angle $i$ changes the velocity along the line of sight by a factor of $\cos(i)$, and the position on the plane of the sky together with the proper motion by a factor of $\sin(i)$. Both of these factors equal each other and $2^{1/2}=0.7071$ for our choice of $45\arcdeg$, canceling the angle factors in this case ($\tan(45\arcdeg)=1$) when the gradients are computed. This explains the special angle and slope exhibited by the out-bending jet velocity centroids in Figure \ref{fig:6}. Hence, the apparent shifts of jet velocity centroids inferred from the PV diagrams give a direct indication of the magnetic acceleration, corrected for the individual inclination angle of the system (see Section \ref{sec:protostellar_candidates}).

\addtocounter{figure}{1}
\begin{figure*}
\figurenum{\arabic{figure}}
\centering
\epsscale{1.1}
\plotone{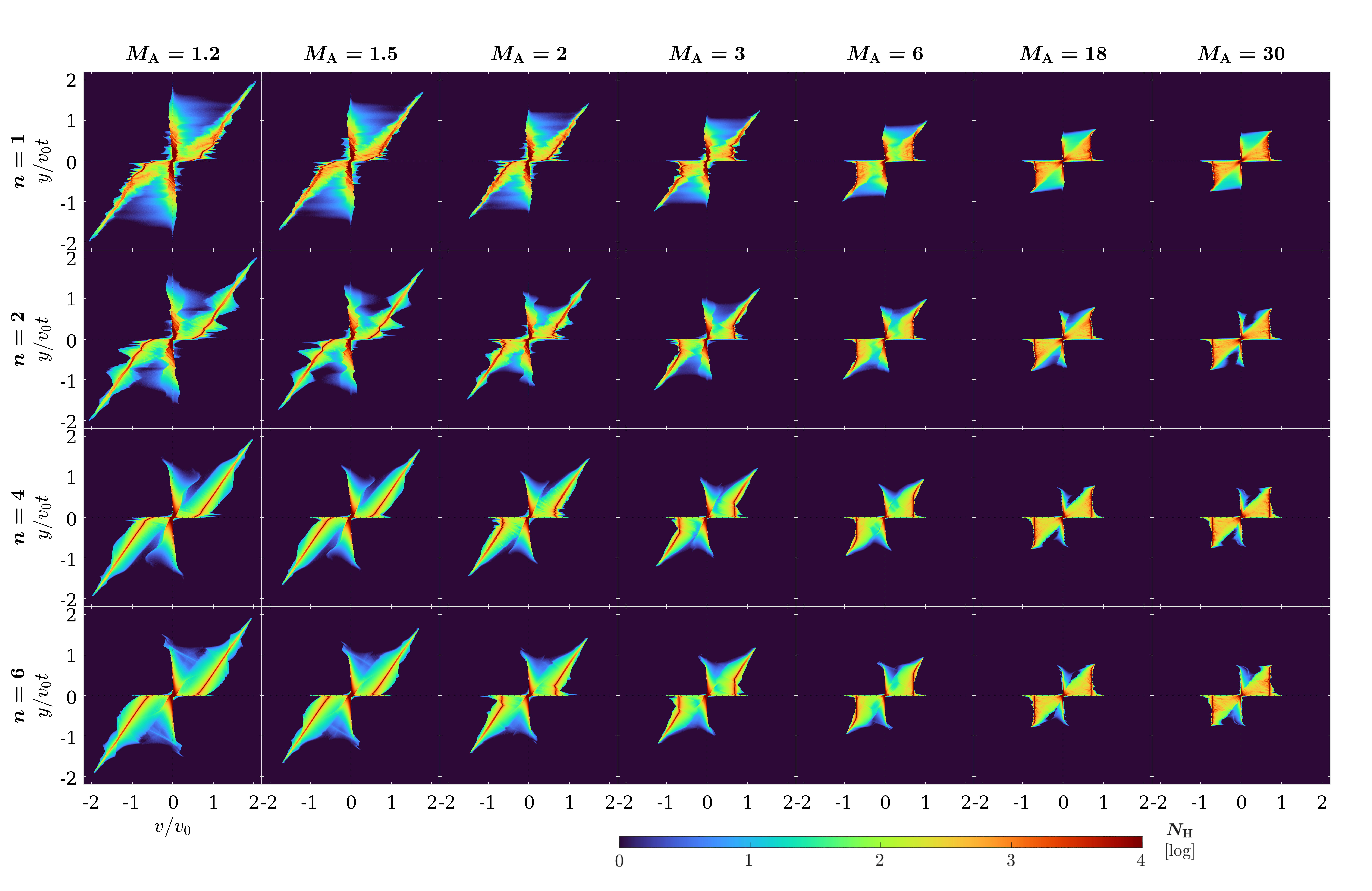}\\
\plotone{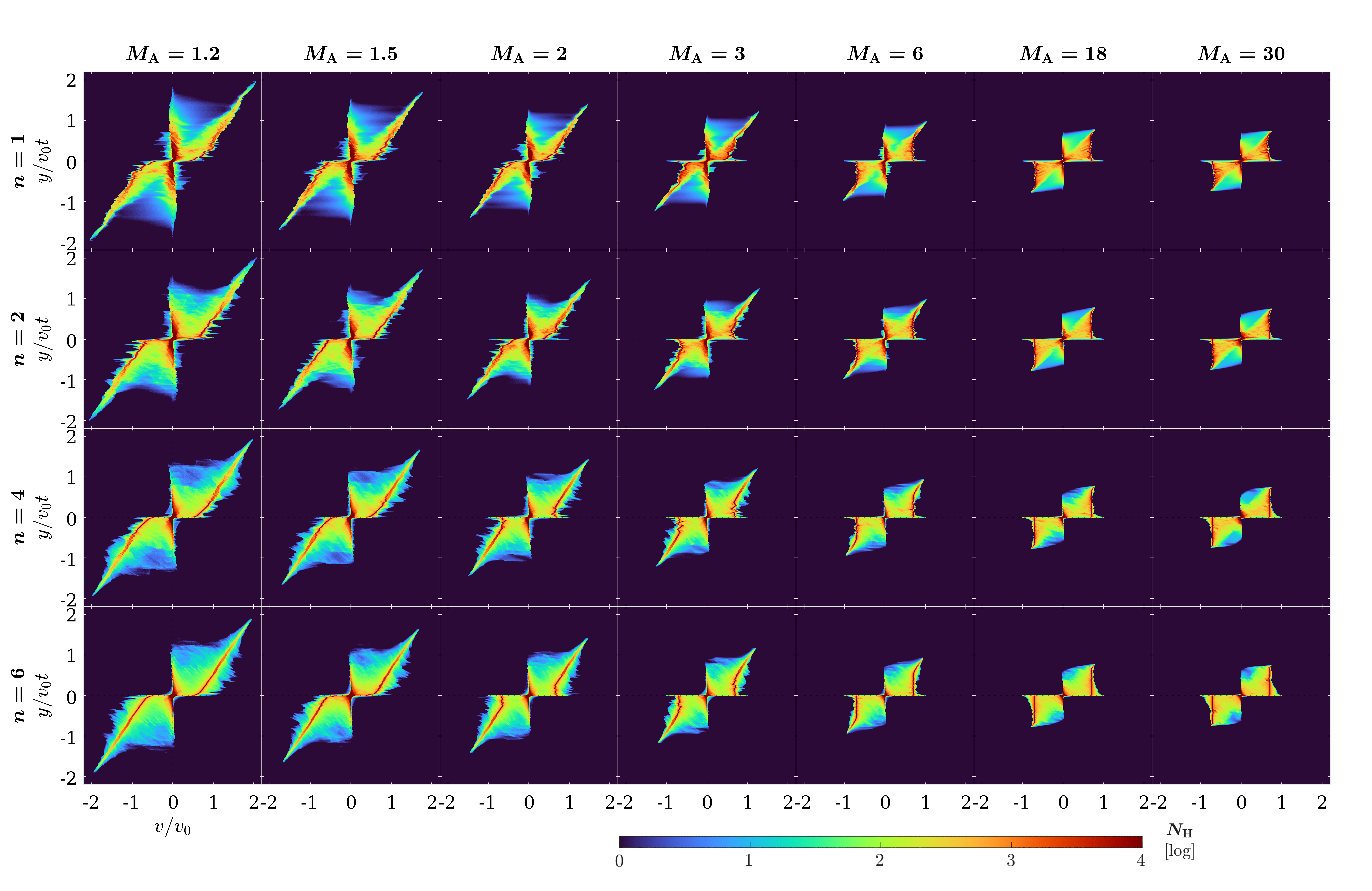}
\epsscale{1.0}
\caption{(a) PV diagrams of column density (rescaled for $\log N_\mathrm{H}$ contrast) produced by winds with $\MA=1.2$, $1.5$, $2$, $3$, $6$, $18$, and $30$ (left to right) for the strongly magnetized $\alphab=1$ (top) and nonmagnetized $\alphab=0$ (bottom) toroids of $n=1$, $2$, $4$, and $6$ (top to bottom) at an inclination angle of $45\arcdeg$. The column density is integrated for material with $v_p>a_\mathrm{ambient}$. The spatial position is convolved with a Gaussian profile of $0.01v_0t$ for smoother appearance as in \citetalias{shang_PII}.}
\label{fig:6}
\end{figure*}

\begin{figure*}
\figurenum{\arabic{figure}}
\centering
\epsscale{1.1}
\plotone{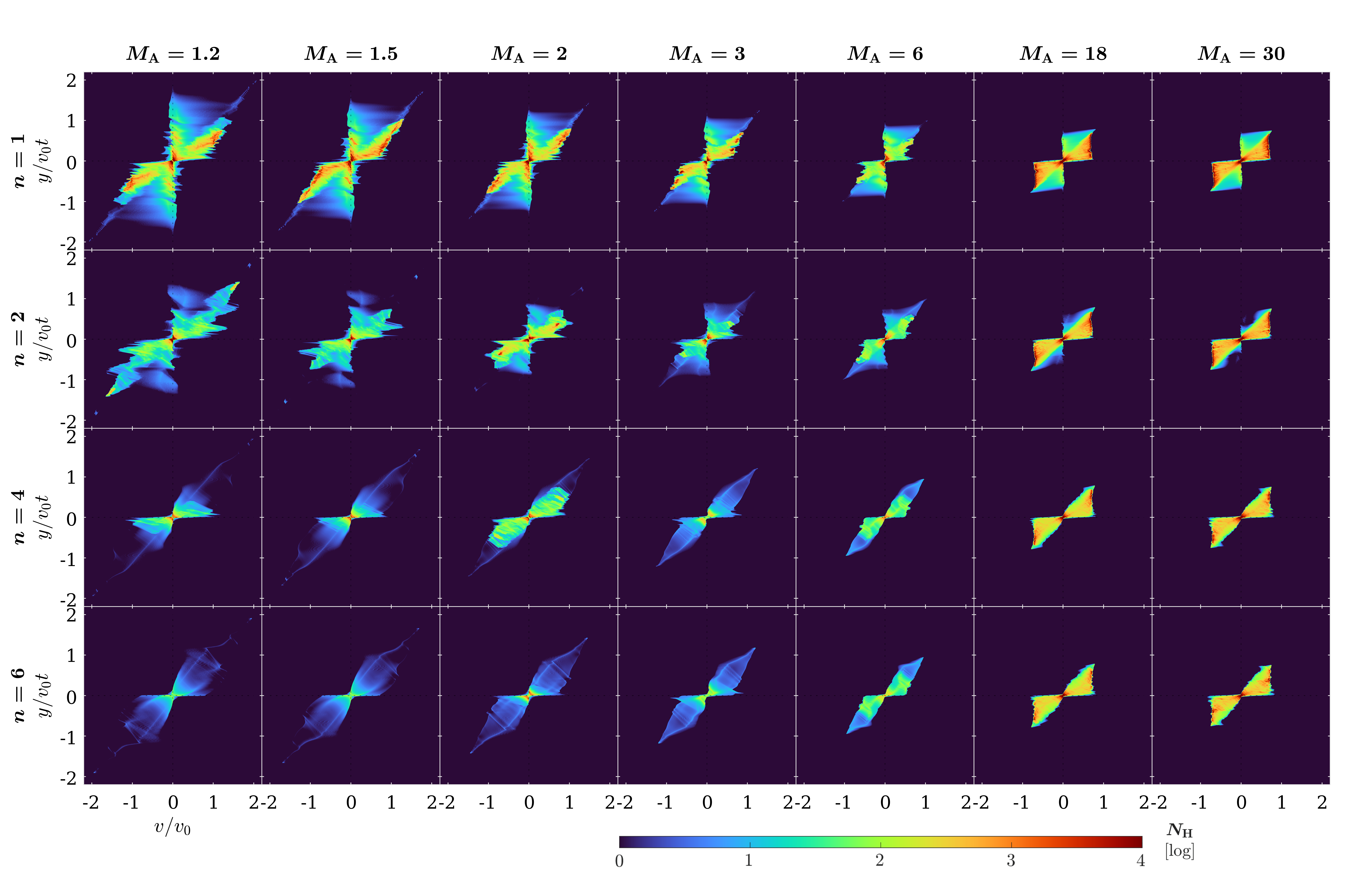}\\
\plotone{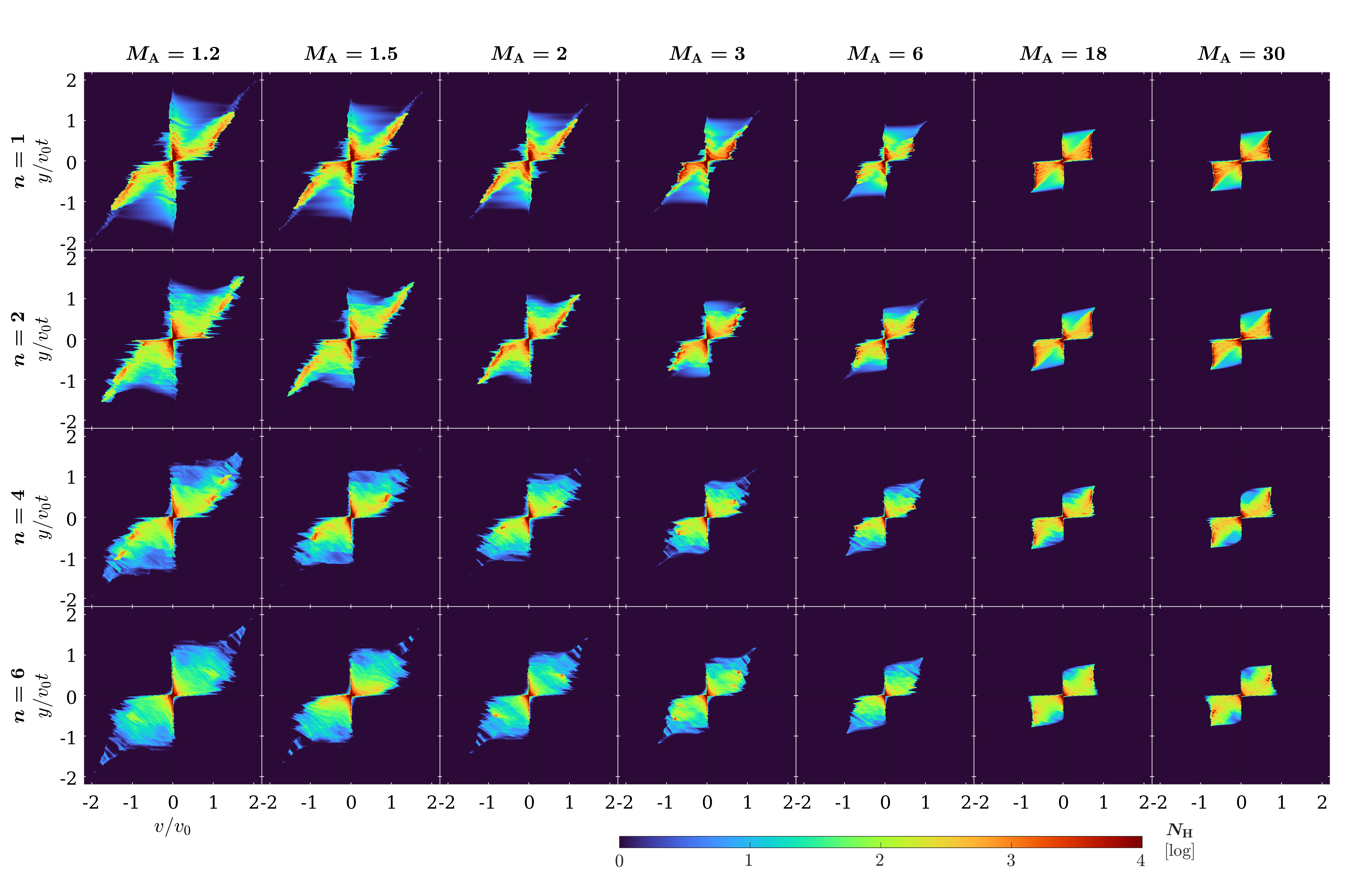}
\epsscale{1.0}
\caption{(b) PV diagrams of column density (rescaled for $\log N_\mathrm{H}$ contrast) as in Figure \ref{fig:6} (a), while the matter is integrated for wind mass fraction $0<f<1$.}
\end{figure*}

This apparent acceleration at large scale is a feature that increases with the strength of the wind magnetization. This effect is to be observed in appropriate systems that can properly manifest the phenomenon. For extremely magnetized systems, this acceleration can increase $v_r$ to $\sim2$--$3$ times the terminal velocities and expand the outflow lobe to $\sim2$--$3$ times the originally expected sizes.  For extremely strong small $\MA$ values, the acceleration zone will dominate most of the jet proper, leaving a very small cavity enclosing the free wind relative to the volume of the outflow bubble.

When the system exhibits such behavior at a moderate $20$--$60$\% level, the effect could otherwise be interpreted as a change of jet activities, especially when combined with intrinsic system episodicities caused by sporadic ejections. It is expected that for moderately to weakly magnetized systems, the effect is relatively weak and confined to a tiny tip or entirely nonexistent. Absence or no detection in these systems suggests magnetization in the underlying wind is likely not strong enough for the apparent acceleration to be visible.

In situations where actual pulses are present, as happens in real systems, similar signatures would apply for the formation of the acceleration zone and each of the pulses after the first pulse. The first acceleration zone should be located farthest downstream of the most extended compressed-wind region filled with pseudopulses. If the nonsteady wind follows the same $1/\varpi^2$ profiles, it will maintain this self-similarity discussed in Sections \ref{sec:method} and \ref{sec:numerical}. The nonsteady wind will be self-shocked and self-similar because the factors cancel out. The nonsteady nature will not affect the first pulse that determines the outermost acceleration zone; however, each of the pulses can establish its own acceleration zone and its compressed-wind region bounded by each new pair of internal reverse-forward shocks. Despite their similarity in physics and structures, each new zone will be interrupted by new pulses and new layers of shocked cavities. Within the interrupted acceleration zones, the velocities may not be able to develop to complete growth factors indicative of the underlying $\MA$ values. Detailed structures of the dynamically large pulsed ejections deeply connected to accretion or other system perturbations will be explored and reported in future publications.

\subsection{Implications}
\label{subsec:implications}

We have demonstrated a potentially efficient acceleration mechanism for an extremely toroidally magnetized jet system with an extended intershock region between the reverse and forward shocks. The shocks are oblique locally and part of a very elongated self-similar bubble. This mechanism may be able to operate in similar extended inter-shock structures formed by configurations meeting the conditions outlined.

The mechanism outlined gives a new angle to understand the observed jet activities. Up to now, most variations in jet velocities inferred from proper motions or from the PV diagrams have been interpreted as the episodicity of the ejections, most likely tied to the underlying accretion. Systematic faster proper motions from knots located farther out have been simply interpreted as faster ejection velocities in the long past. Given the physics demonstrated in this work, such systems may be our candidates for manifesting the phenomena of apparent acceleration. However, due to a potential lack of shocking pulses in the top portion of the extended acceleration region, the identification of a proper detectable radiative mechanism may be needed. What systems to search for the signatures shall be our next challenge in future works.

We observe the jets formed and sustained by very strong toroidal magnetic fields remain stable for the large physical scales, despite the presence of KHI at the interface with the ambient medium and within the initial compressed and shocked regions. The top (tip) portion of the acceleration zone, however, appears to be KHI-free and stable. Vorticity seems to be suppressed, which has prohibited the KHI\@. This finding can help us understand how parsec-long jet systems can remain stable without being disrupted or fragmented by the instabilities.

M87 is known to have a helical field \citep{pasetto2021}, which is consistent with Faraday depolarization at the projected axis and the edges of the conical jet. It also has an extended region with KHI filamentary patterns \citep{lobanov2003,hardee2011}, which coincides with the twisting of the toroidal fields \citep{blandford_araa} within the conical jet. The M87 jet has an extended lobe-like far jet beyond the conical jet in the VLA image in Figure 1 of \citet{pasetto2021}, which strongly suggests the possibility of an underlying bubble structure. The knots in the conical jet and at least the beginning of the far jet (knot A) have been proposed to be produced by KHI, again sharing support for our present framework in the compressed-wind region in the regime dominated by magnetic pressure.
Within this Newtonian acceleration picture, partial spending of (e.g.) $\textthreequarters$ of the available specific magnetic enthalpy (leaving $\textonequarter$ of $h_i\approx h_0$) would correspond to an intermediate speed $v_m/v_0=1+1/\MA<v_f/v_0=1+2/\MA$,
suggestive of a situation of transition between jet regions.
The kind of acceleration mechanism and regions formed with the extremely magnetized bubble by toroidal magnetic configuration can be relevant to M87 and perhaps other black hole jet systems dominated by a toroidal magnetic field. 
The correlation with the measured Faraday depolarization and helical field structure may not be a coincidence.

\subsection{Protostellar Candidate Systems}
\label{sec:protostellar_candidates}
Similar systems of very magnetized bubbles relevant to the regime addressed in this work already exist in the literature. 

\subsubsection{HH 212}

HH 212 is recognized as a candidate system exhibiting magnetized bubble features, as demonstrated in Section 8 of \citetalias{shang2020} and Section 9.5 of \citetalias{shang_PII}\@. It demonstrates the signature multicavities of nested shell-like structures and velocity profiles.

Apparent long-range acceleration has also been observed and reported in HH 212 based on historical proper-motion measurements \citep{lee2022}, which is otherwise deemed difficult for most traditional magnetocentrifugal-launching models of free winds. Figure 2 of \citet{lee2022} shows the proper motions of SiO knots from $\sim100$ to $1000\au$ and much larger distances (up to $\sim 16000\au$) along the jet axis \citep{claussen1998,lee2015,reipurth2019}. There, an apparent acceleration is seen over the distance from $50\au$ to $3000\au$, with the velocity increasing from $\sim50$--$60\kms$ at $40$--$50\au$ to $\sim140\kms$ at $3000\au$, approximately a factor of $\sim 3$ increase over a scale of more than two orders of magnitude (Figure 2 of \citet{lee2022}). This increase by a factor of $\sim 3$ from $50\kms$ to $140\kms$ at $3000\au$ is beyond the increase predicted by the asymptotics in \citet{shu1995}.

HH 212 is not only an excellent testbed of the multicavities as predicted by our framework \citepalias{shang2020,shang_PII} but also a lab for multiepoch large ejections of multiscale nested self-similar bubble structures. For a better understanding, the first segment of the apparent large shell up to the scale of $\sim 1200\au$, is followed by an earlier ejected shell of $\sim 16000\au$. The historical record of the proper motions should be grouped according to their coeval nested shell structures formed with different histories of bubble formation.

In the recently established shell as part of the self-similar bubble of $1200\au$, we obtain the velocity along the line of sight (on the order of $10\kms$) from the PV and as proper motion (on the order of $100\kms$) for knots from both sides located at positions around $500$--$1000\au$ from Figures 2 and 5 in \citet{lee2022}. For the edge-on system with $i\approx84\arcdeg$ in HH 212, we obtain the gradients based on proper motions on three accelerating knots (located respectively at $y\approx590$, $710$, and $810\au$). The gradients are
shown to be constant and compatible with a dynamical time $t\approx20\yr$ obtained from Equation (\ref{eqn:acc:e30}). A simple visual inspection of the PV diagram shows a straight line connecting the velocity centroids, giving the same constant gradient as for the proper motion once we adjust for the viewing angle $i$ with a factor $\approx10\approx\tan(84\arcdeg)$, an intuitive result that is confirmed in the more careful analysis, leading to the same gradient and dynamical time of $t\approx20\yr$.

The segment of three accelerating knots with the constant gradient on the scale of $500$--$1200\au$ appears to follow a segment of knots oscillating around a constant centroid of $\sim 55\kms$. This region corresponds to the compressed region within the $1200\au$ bubble cavity, where the apparent acceleration occurs. Out to an even larger scale of $\sim 16000\au$, \citet{lee2022} also find that velocity further away remains roughly constant, which in our interpretation belongs to the compressed-wind region of the bubble cavity established by an earlier event. These large nested cavities appear self-similar to each other, and they belong to bubble cavities formed at different epochs of large ejections. This is consistent with the large-scale pseudopulses filling the previous generation of compressed-wind region outward to the previous terminal shock (see the discussion in \citetalias{shang_PII}), and that real large pulses by ejection events establish a new bubble nested within the previous one. We will follow up with further theoretical explorations in a later work.

\subsubsection{HH 211}

The Class 0 system HH 211 is one of the youngest sources driving a highly collimated molecular jet and an outflow \citep{gueth1999,hirano2006,palau2006,lee2007b}. It is also the only known Class 0 jet surrounded by a toroidal configuration of a magnetic field established by SiO line polarization \citep{lee2018_hh211_GK} at the time of writing. Hence, it is a desirable testbed candidate for our theoretical configurations with a confirmed magnetic field structure.

The signature of this apparent second acceleration can be observed and extracted from existing published data from the Submillimeter Array (SMA)\@. In the PV diagrams of CO and SiO in \citet{hirano2006}, \citet{palau2006}, and \citet{lee2007b}, the redshifted emission traces a shift of the velocity centroid increasing from $25 \kms$ to $35 \kms$ beyond $\sim10\arcsec$ in  \sio{5}{4}, \sio{8}{7}, \co{2}{1}, and \co{3}{2}. In Figure 3 of \citet{hirano2006}, the shift is continuous and follows a straight line in \sio{5}{4}. In Figure 3 of \citet{palau2006} and Figure 5 of \cite{lee2007b}, \sio{8}{7} also traces the same shifted velocity centroid. \co{3}{2} in Figure 5 of \citet{lee2007b} traces a straight line following the trend in \sio{8}{7} and \sio{5}{4}. These emission knots are identified between the SiO and H$_2$ knots RK5 to RK7 that are $\sim3\farcs5$ apart, with a rough increase of $\sim4.5\kms$. Combined with the inclination angle $i\approx80\arcdeg$ \citep{hirano2006} and a distance of $321\pc$ \citep{ortiz-leon2018}, an apparent acceleration is clearly shown to have a constant slope, compatible with the gradient of Equation (\ref{eqn:acc:e30}) with a dynamical time of approximately $110\yr$. We await further analysis of ALMA data in multiple SiO and CO transitions in future works.

\subsubsection{Radio Continuum Jets}

Radio jets from more massive protostellar systems show increasing velocities measured via proper motions of knots going away from the source. In recent observations, several massive protostars show indications of jets and wide-angle wind cavities. This suggests that some similar jet-driving mechanisms may operate deep down at the base of these sources.

\citet{rodriguez-kamenetzky2022} reported a resolved deeply embedded collimation zone ($\lesssim100\au$) at $\sim15\au$ resolution of an intermediate-mass Class 0 protostar \citep[$\sim3\,\msun$ and 100 $\lsun$;][]{hull2016} in Serpens, known as the SMM1. Its radio jet consists of a central elongated thermal source (C) and two external lobes (NW and SE, $3000\au$ apart) \citep[e.g.,][]{rodriguez1989,curiel1993,rodriguez-kamenetzky2016}. NW and SE have been known to travel at tangential velocities $\gtrsim 200$--$300\kms$ in opposite directions \citep{rodriguez1989,rodriguez-kamenetzky2016}, and the internal knots tend to have velocities $\sim 100\kms$. Periodic velocity variations due to tidal interactions, a binary companion with an eccentric orbit, or a past FU-Ori outburst event have been proposed as causes of such velocity trend \citep[e.g.,][]{rodriguez-kamenetzky2022}. However, a factor of 2 velocity increase can be observed on each side separately: SE\_N/(S1 \& S2), and NW/(N3 \& N4), using the information on the knots compiled in Table 3 of \citet{rodriguez-kamenetzky2022}. This implies a highly magnetized flow with an equivalent $\MA\sim 2$ that can be inferred using the formulation derived in Section \ref{subsec:maxvel} with Equation (\ref{eqn:acc:f26}). 

SMM1 is one of the few Class 0 sources with radio jets and has molecular cavities. It also has nested shell structures from \co{2}{1} \citep{hull2016,tychoniec2019,tychoniec2021}, in widening cavity layers and decreasing velocity components (EHV, fast, slow) surrounding the radio jet observed by e-MERLIN \citep{rodriguez-kamenetzky2022}. This is consistent with the kinematic and morphological picture of a wind-driven elongated magnetized bubble developed in \citetalias{shang2020} and \citetalias{shang_PII}. On the physical scale of $\lesssim 100 \au$, 
a collimation zone is resolved with an ionized stream at $60\au$ within a narrow cavity of $\sim 28 \au$ near the source. This structure is likely the RS cavity enclosing the primary free X-wind (or an X-wind-like inner disk wind) launched from $\sim 0.4\au$. Our current work adds additional information on velocity ratios based on the second acceleration and an inferred strength of wind magnetization of $\MA\sim 2$ from our framework. The higher luminosity of $100\lsun$ for $3\msun$ possibly helps with the ionization and illumination of the radio knots. Similar systems would help reveal more massive protostellar systems and their connections to magnetized jets through the proper-motion measurements of knots located far from the embedded sources through a different observational approach using molecular lines.

\section{Summary}
\label{sec:summary}

We advance a new mechanism of magnetic acceleration within a self-similar highly elongated bubble driven by a strongly toroidally magnetized super-fast wind resulting from the magnetic interplay with the ambient medium. This bubble features a pinched tip along the jet axis generated by a cylindrically stratified density and magnetic field profiles.

This dynamical acceleration operates far beyond the initial launch and acceleration of the usual primary magnetocentrifugal mechanism and the free asymptotic state. The system enters a nonsteady dynamical phase, in which the toroidal wind magnetic field is converted into the increase of radial velocity toward the upper tip of the elongated bubble. The acceleration can maintain a shape-independent constant slope, reaching an $\MA$-dependent maximum velocity, determined at the initial ejection. An ideal system will demonstrate the correlation between the decrease of toroidal magnetic field strength and the increase of velocity toward the terminal shock far beyond the reverse shock.

Predicted observational signatures may be identifiable from systematically stronger and faster flow/jet velocity in the past, or otherwise unrecognizable in cases with a weaker jet magnetization. 
Three protostellar sources are discussed and show features compatible with the present mechanism. The constant velocity gradient of HH 211 and HH 212, a prediction of this mechanism, has been observed in PV for both sources and in proper motion for HH 212. The acceleration to larger velocities (by a factor of 2) in SMM1 and HH 212 can fit the terminal velocity of this mechanism. Additionally, the source M87 has basic physics potentially conducive to this mechanism (toroidally magnetized wind and jet), and we tentatively propose that it can contain a relativistic analog of the presented Newtonian mechanism.

\begin{acknowledgments}

The authors would like to thank the anonymous referee whose comments and suggestions have significantly improved the presentation of this work. The authors acknowledge grant support for the CHARMS group under Theory from the Institute of Astronomy and Astrophysics, Academia Sinica (ASIAA), and the National Science and Technology Council (NSTC) in Taiwan through grants 110-2112-M-001-019-
and 111-2112-M-001-074-. The authors acknowledge the access to high-performance facilities (TIARA cluster and storage) in ASIAA, and thank the National Center for High-performance Computing (NCHC) of National Applied Research Laboratories (NARLabs) in Taiwan for providing computational and storage resources. This work utilized tools (Zeus-TW, \Synline{}) developed and maintained by the CHARMS group.
This research has made use of SAO/NASA Astrophysics Data System.

\end{acknowledgments}

\software{\Synline{} \citep{shang_PII} CHARMS (ASIAA), Zeus-TW \citep{krasnopolsky2010} CHARMS (ASIAA), MATLAB, Matplotlib \citep{hunter2007}.}

\newpage

\bibliographystyle{aasjournal}
\bibliography{Outflows}

\appendix

\section{Simulation Setup}
\label{sec:setup}

For the calculations carried out in this work, we adopt numerical values and notations consistent with the numerical simulations in \citetalias{shang2020} and \citetalias{shang_PII}. The formulation, expressed in Lorentz--Heaviside (LH) cgs units for the magnetic field vector $\bb$, is consistent with the Zeus-TW code \citep{krasnopolsky2010}.
The constructed free wind $\rho_\mathrm{w}$ and the tapered toroid $\rho_\mathrm{a}$ are
\begin{eqnarray}
r^2 \rho_\mathrm{w} &= D_0 / \sin^2\theta, \, \label{eqn:rhowbar} \\
r^2 \rho_\mathrm{a} &= (\lfrac{a_\mathrm{a}^2}{2\pi G}) R(\theta) + D^S\ \label{eqn:rhoabar},
\end{eqnarray}
where $R(\theta)$ (a dimensionless density function; and also $\phi(\theta)$, a dimensionless flux function) is found from the solutions of the toroids \citep{li1996b,allen2003}. The initial magnetic field in the toroids is purely poloidal, and it has the value $\bb=\alphab\bb^\mathrm{T}(r,\theta)$, where $\alphab$ is a scaling constant, multiplying the toroid field $\bb^\mathrm{T}$ computed using $\phi(\theta)$. The density constants adopted for the wind $D_0$, the tapered toroid $D^S$, the density scale factor $\Tilde{D}$, and the toroid magnetic scale $\alphab$ have the numerical values listed in Table \ref{tab:quantities} below.

A two-temperature equation of state \citep{wang2015} is built in for the pressure $p$ and sound speed $a$, depending on the wind mass fraction $f=\rho_\mathrm{wind}/\rho$, and the ambient sound speed $a_\mathrm{ambient}=0.2\kms$ and the wind sound speed $a_\mathrm{wind}=0.6\kms$:
\begin{equation}
 \label{eqn:eos}
 \begin{split}
    p = &\ a^2\rho\\
    a^2 = &\ f a_{\mathrm{wind}}^2 + (1-f) a_{\mathrm{ambient}}^2\\
 \end{split}
\end{equation}
Most of this article focuses on the cold and essentially unmixed wind regions of the outflow, such that $f\approx1$, $a\approx{a_{\mathrm{wind}}}$, and $\bb_p$ and $p$ are minor or negligible terms in the force equations.

\begin{deluxetable}{ccccc}
\tablecaption{Quantities Defined in Papers I, II, and This Work}\label{tab:quantities}
\tabletypesize{\scriptsize}
\tablewidth{\textwidth}
\tablecolumns{5}
\tablehead{Symbol & Description & Definition & Adopted Value(s) & Reference}
\startdata
\multicolumn{5}{c}{Coordinate Systems} \\
\hline
$(r,\theta,\phi)$ & spherical polar coordinates & & & \\
$(\varpi,z,\phi)$ & cylindrical polar coordinates & & & \\
\hline
\multicolumn{5}{c}{Vector Quantities} \\
\hline
$\vv$  & flow velocity & & & I, II \\ 
$\bb$  & magnetic field in LH units & & & I, II \\
$\jv$  & current density in modified LH units & $\jv\equiv\nabla\times\bb$ & $\jv_p=(\lfrac{1}{\varpi})\vhat\phi\times\nabla\Cfunc$ & I, II \\
$\fv_c$  & specific magnetic force & $\jv\times\bb$ & & I, II \\
$\omv$ & vorticity & $\nabla\times\vv$ & & I, II \\
\hline
\multicolumn{5}{c}{Scalar Quantities} \\
\hline
$\rho$ & gas density & & & I, II \\
$f$    & wind mass fraction & $\rho_\mathrm{wind}/\rho$ & & I, II \\
$a_\mathrm{wind}$    & wind sound speed & & $0.6\kms$ & I, II \\
$a_\mathrm{ambient}$    & ambient sound speed & & $0.2\kms$ & I, II \\
$a$    & local sound speed & $\left[f a_{\mathrm{wind}}^2 + (1-f) a_{\mathrm{ambient}}^2\right]^{1/2}$ & & I, II \\
$p$    & gas pressure & $a^2\rho$ & & I, II \\ 
$\mathcal{M}$ & mass per steradian & & & I, II \\
$\dot{\mathcal{M}}_\mathrm{wind}$ & wind mass-loss rate & & $=3.6415\times10^{-6}\solarmassyr$ & I, II \\
$\Phi$ & magnetic flux function & & & I, II \\
$\tirho$ & modified density & $\rho\varpi^2$ & & this work \\
$\tib$   & modified toroidal field & $b_\phi\varpi$ & & this work \\
$\mu$  & magnetic-to-mass ratio & $\tib/\tirho$ & & this work \\
$\Cfunc$ & current function & $-b_\phi\varpi$ & & I, II \\
$\Lfunc$ & force function & $\Cfunc^2/2$ & & I, II \\
$h$    & specific magnetic enthalpy & $\mu\tib$ & & this work \\
$H$    & \parbox{3cm}{Bernoulli constant in the acceleration zone} & $\lfrac{v_r^2}{2}+h-\lfrac{r^2}{(3t^2)}$ & & this work \\
$D_t$  & material derivative & $\partial_t+\vv\cdot\nabla\approx\partial_t+v_r\partial_r$ & & this work \\
$x$    & self-similar coordinates & $r/v_0t$ & & this work \\
$i$    & \begin{tabular}{c} inclination angle, between\\ the line of sight and outflow axis\end{tabular} & & see text & II \\
\hline
\multicolumn{5}{c}{Parameters} \\
\hline
$n$    & opening of the initial toroid & & $1$, $2$, $4$, $6$ & I, II \\
$\alphab$ & ambient poloidal field strength factor & & $0$, $0.1$, $1$ & I, II \\
$\alpha_\rho$ & initial toroid density factor & & $0$ (without toroid), $1$ (with toroid) & I, II \\
$v_\mathrm{w}$ & radial wind velocity & & & I, II \\
$v_s$  & radial shell velocity & & & I, II \\
$\rin$ & position of the inner radial boundary & & $1.5\au$ & I, II \\
$\rout$ & position of the outer radial boundary & & $10^5\au$ & I, II \\
$G$  & Newton's constant & & $6.67428\times10^{-8} \cm^3\gram^{-1}\second^{-2}$ & I, II \\
\hline
\multicolumn{5}{c}{Dimensionless Angular Functions for the Initial Conditions of Winds and Toroids} \\
\hline
$R(\theta)$ or $Q(\theta)$ & toroid density distribution function & & & I, II \\
$\phi(\theta)$ & toroid poloidal magnetic flux function & & & I, II \\
$P(\theta)$ & momentum input function of the wind & & & I, II \\
$\mathcal{B}(\theta)$ & bipolarity function of the outflow & & & I, II \\
\hline
\multicolumn{5}{c}{Initial and Boundary Conditions} \\
\hline
$v_0$  & wind velocity from the inner boundary & & $50$, $100\kms$ & I, II \\
$b_0$  & wind toroidal magnetic field constant & & $v_0\sqrt{D_0}/\MA$ & I, II \\
$D_0$  & wind density constant & & $2.0025\times10^{11}\gram\cm^{-1} \times(v_0/100\kms)$ & I, II \\
$D^S$  & tapering density factor for the toroid & & $2.25\times10^{11}\gram\cm^{-1}$ (or zero) & I, II \\
$\Tilde{D}$ & density scale factor for the toroid & $\lfrac{a^2_\mathrm{a}}{2\pi G}$ & $=9.5384\times10^{14}\gram\cm^{-1}$ & I, II \\
$\MA$  & Alfv\'enic Mach number of the wind & $v_0\sqrt{D_0}/b_0$ & see text & I, II, this work \\
$\mu_0$ & magnetic-to-mass ratio in the free wind & $b_0/D_0$ & & this work \\ 
$h_0$  & specific magnetic enthalpy in the free wind & $v_0^2/\MA^2$ & & this work \\
\hline
\multicolumn{5}{c}{Subscripts and Superscripts} \\
\hline
$q_0$  & quantities at the inner radial boundary & & & I, II \\
$q_p$  & quantities in the poloidal component & & & I, II \\
$q_\mathrm{w}$ or $q_\mathrm{wind}$ & quantities of the wind & & & I, II \\ 
$q_\mathrm{a}$ or $q_\mathrm{ambient}$ & quantities of the ambient & & & I, II \\ 
$q_\mathrm{MC}$ or $q_\mathrm{S}$ & quantities for the momentum-conserving thin shell & & & II \\ 
$q_\mathrm{RS}$ & quantities for the reverse shock & & & II \\ 
$q_\mathrm{FS}$ & quantities for the forward shock & & & II \\ 
$q\ib$  & quantities at the start of the acceleration zone & & & this work \\
$q_f$ & quantities at the end of the acceleration zone & & & this work \\
$q^\mathrm{T}$ & quantities of the toroid & & & I, II \\
$q^S$ & quantities for tapering the toroid & & & I, II \\
\hline
\enddata
\end{deluxetable}

\newpage
\section{The Tips}
\label{sec:tips}

In \citetalias{shang2020}, we noted that a peculiar pointy-nose type of structure can arise in the (small) tip region, that the flow is nearly radial ($v_r\gg v_\theta$), and that values of $b_p$ are nearly zero.
These are similar to the assumptions adopted for the free wind. The transport equations for mass $\rho$ and $b_\phi$ are similar. 

Hence, the radial momentum equation can take a simplified form, ignoring terms in $v_\theta$ and $v_\phi$, as in a nearly radial wind:
\begin{equation}
    \begin{split}
    \frac{D v_r}{Dt}&=
    \partial_t v_r+v_r\partial_r v_r=(v_r-r/t)\partial_r v_r\\
    &=
    -\frac{\partial_r p}{\rho}-\frac{\Cfunc\partial_r(\Cfunc)}{\rho\varpi^2}\ .
    \end{split}
\end{equation}
Ignoring the gas pressure term, the force equation is simplified to
\begin{equation}
\label{eqn:eqn53_paperI}
\begin{split}
    (v_r-r/t)\partial_r v_r
    &=\mu\partial_r\Cfunc\approx\partial_r(\mu\Cfunc)\ .
\end{split}  
\end{equation}
Inside the tip region, the magnetic force term is able to smoothly accelerate $v_r$, while $b_\phi$ smoothly decreases along the $z$ direction. The observation that the tip region follows the expected behavior of the simplified force equation 
motivates the generalization in Section \ref{sec:method}. The behavior of $\mu$ is explored in more detail in Appendix \ref{sec:mu}.

\section{Properties of \texorpdfstring{$\mu$}{μ}}
\label{sec:mu}

\begin{figure*}
\centering
\epsscale{1.1}
\plottwo{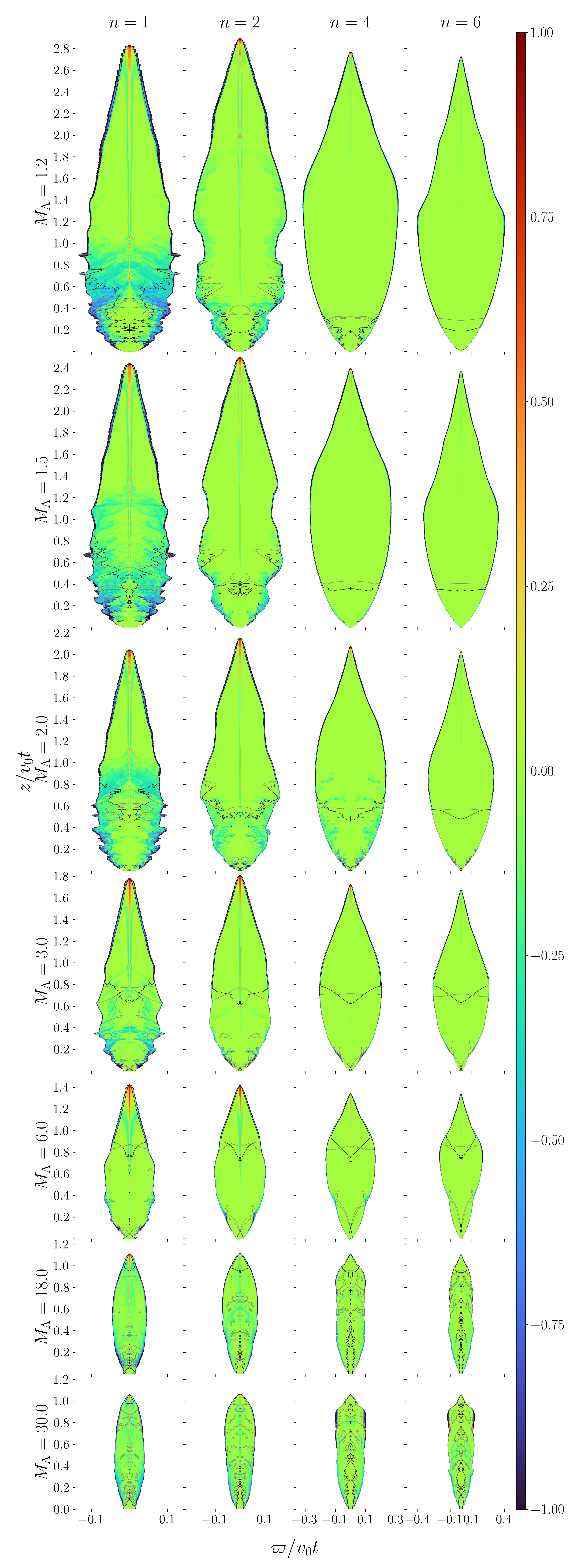}{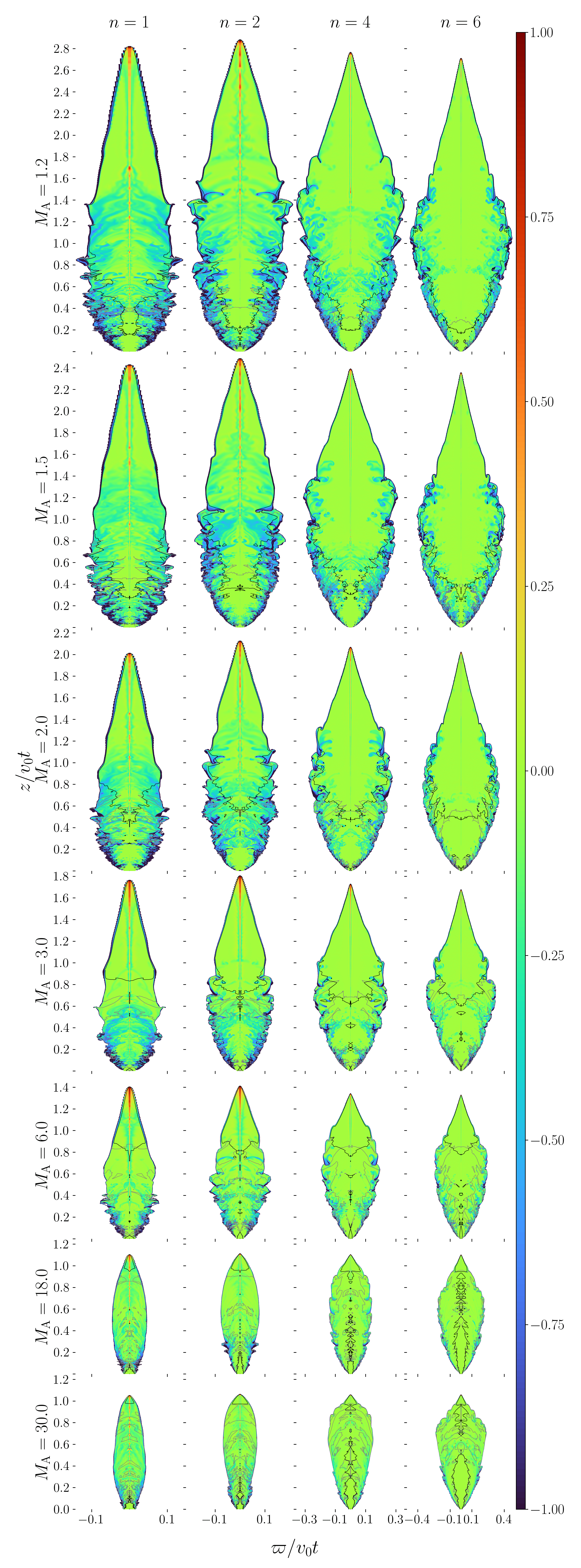}
\epsscale{1.0}
\caption{The 2D spatial profiles of the quantity $\log_{10}\left(\mu/\mu_0\right)$ for winds of $\MA=1.2$, $1.5$, $2$, $3$, $6$, $18$, and $30$ with $n=1$, $2$, $4$, and $6$ toroids and ambient poloidal field strengths of $\alphab=1$ (left) and $\alphab=0$ (right), in which $\mu_0 \equiv v_0/\MA\sqrt{D_0}$. The black contours are loci of $v_r/v_0=1$, and the brown contours are loci of $-\Cfunc/b_0=1$. The spatial axes are labeled in units of $v_0t$. The horizontal $\varpi$ axes have been exaggerated by factors of $3.5$, $2.3$, $1.4$, and $1.0$ for $n=1$, $2$, $4$, and $6$, respectively.}
    \label{fig:7}
\end{figure*}

In this appendix we study in some detail the quantity $\mu$, defined in Equation (\ref{eqn:mu}), and utilize it to find some results about magnetism in the self-similar regime of this flow.
We first investigate the variations of $\mu=\tib/\tirho$ by computing the total derivative quantity $D_t\mu$.
Using the usual formula for the derivative of a ratio, 
\begin{equation}
    \label{eqn:acn:e14_1}
    D_t\mu=\tirho^{-2}(\tirho D_t\tib-\tib D_t\tirho)\ .
\end{equation}
Using now Equations (\ref{eqn:acc:e3tilde})--(\ref{eqn:acc:e4tilde}),
\begin{equation}
    \label{eqn:acn:e14_2}
    D_t\mu=\tirho^{-2}[\tirho(-\tib\partial_r v_r)-\tib(-\tirho\partial_r v_r)]
\end{equation}
the last two terms of which cancel to zero, resulting in
\begin{equation}
    \label{eqn:acn:e14}
D_t\mu\approx0
\end{equation}
in a region for which the approximate Equations (\ref{eqn:acc:e3tilde})--(\ref{eqn:acc:e4tilde}) can be applied.
Figure \ref{fig:7} shows the variations of $\mu$ in space as found in numerical computation.
We can find that consistent with Equation (\ref{eqn:acn:e14}) the quantity $\mu$ is nearly constant in most of the smooth regions we consider.
That includes the self-similar laminar acceleration regions where $D_t\mu\approx v_E\partial_r\mu\approx0$.
However, in those parts of the compressed regions where a nonlaminar $v_\theta$ is present due to interplay and vorticity generation, $\mu$ is a rapidly varying function tracing the pseudopulses driven by magnetic forces as shown in Figure \ref{fig:7}.

We utilize now these properties of $\mu$ as a tool to investigate the variations of the magnetic field within the acceleration region.
Substituting $\tirho$ and $\tib$ into and rearranging terms in Equations (\ref{eqn:acc:e1})--(\ref{eqn:acc:e4}), 
and making use of the property of Equation\ (\ref{eqn:acc:e20}), we obtain
\begin{equation}
\label{eqn:acc:e21}
    (v_r-\frac{r}{t})\partial_r v_r+\mu\partial_r\tib=0
\end{equation}
\begin{equation}
\label{eqn:acc:e22}
    (v_r-\frac{r}{t})\partial_r\tib+\tib\partial_r v_r=0\ .
\end{equation}

We multiply Equation\ (\ref{eqn:acc:e21}) by $(v_r-r/t)$, and multiply Equation\ (\ref{eqn:acc:e22}) by $\mu$, reorder the terms, then take the difference of the two resulting equations:
\begin{equation}
\label{eqn:acc:e26}
\left[\mu\tib-(v_r-\frac{r}{t})^2\right](\partial_r v_r)=0\ .
\end{equation}
This result can be fulfilled in two manners: either by having its second factor $\partial_r v_r$ equal to zero as it can happen in the free wind, or by fulfilling Equation (\ref{eqn:acc:e27}) in the main text, thus linking magnetic properties in its left-hand side to hydrodynamic properties in its right-hand side, as it can happen in the acceleration regions which are the focus of this work.

\section{The Acceleration Zone}
\label{sec:accel_zone}

Equation (\ref{eqn:acc:e27}) implies that at this point with $h_f=0$ we also have that $v_f=r_f/t$ and therefore [Equation (\ref{eqn:acc:f11})] $H_f=(1/2-1/3)v_f^2=v_f^2/6$.
The beginning and end of the acceleration region are connected by the $2/3t$ slope, $v_f-v_0=(\lfrac{2}{3t})(r_f-r\ib)$,
rearranged to obtain $r\ib$,
\begin{equation}
    \label{eqn:acc:f17}
    \frac{r\ib}{t}=\frac{1}{2}(3v_0-v_f)\ .
\end{equation}

Substituting $r\ib$ into Equation (\ref{eqn:acc:f10}), and utilizing $H\ib=H_f=v_f^2/6$, leads, after some algebra, to
\begin{equation}
    \label{eqn:acc:f23}
    (v_f-v_0)^2=4h\ib,
\end{equation}
then $v_f-v_0=2{h\ib}^{0.5}$.

\section{Density and Column Density Profiles}

\begin{figure*}
\centering
\epsscale{1.1}
\plottwo{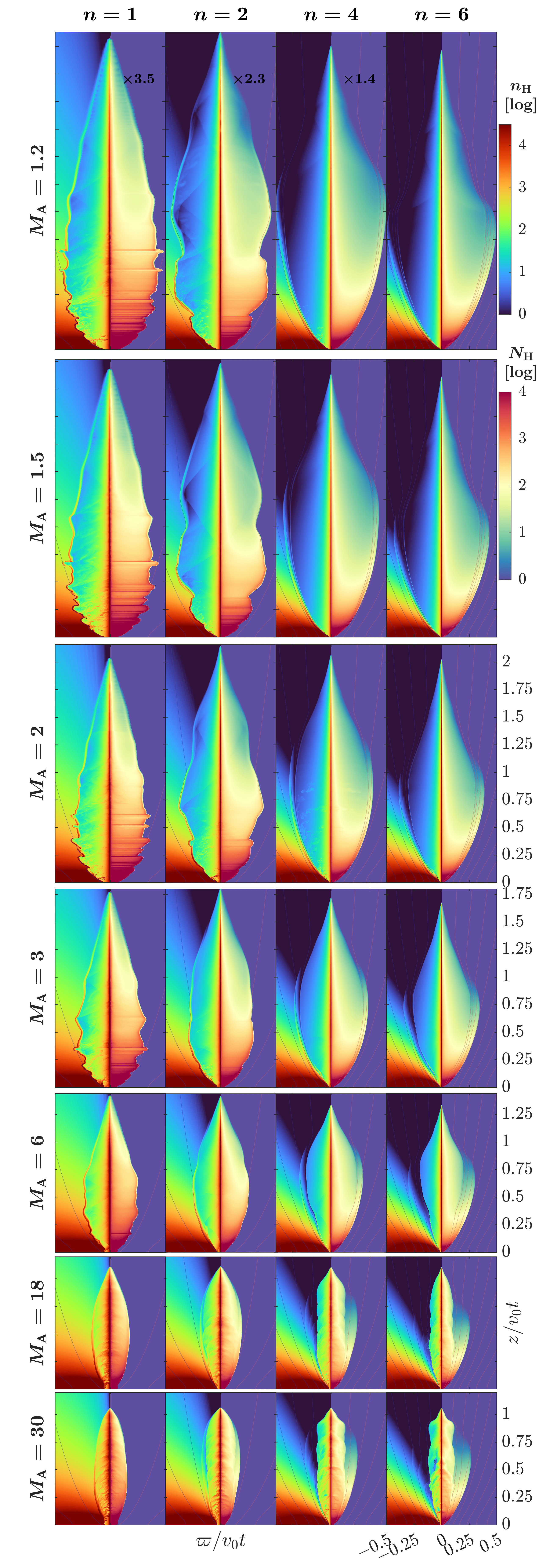}{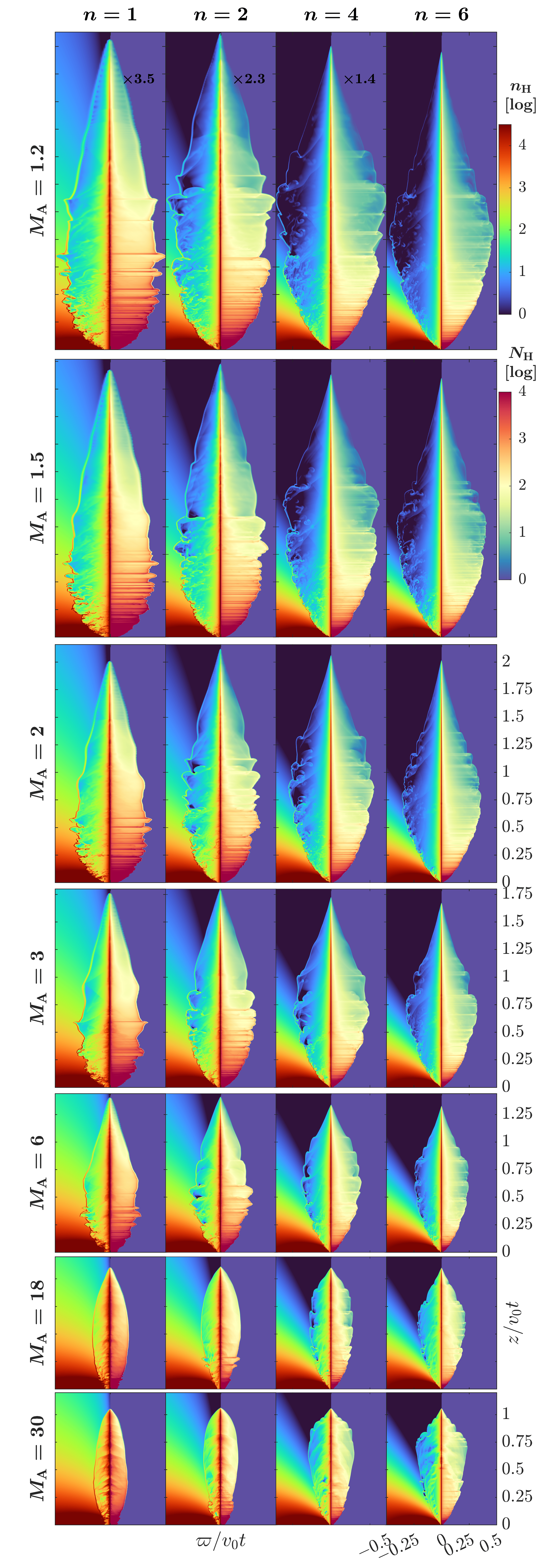}
\epsscale{1.0}
\caption{The 2D spatial profiles of number density ($n_\mathrm{H}$, the left halves) and column density ($N_\mathrm{H}$, the right halves) for winds of $\MA=1.2$, $1.5$, $2$, $3$, $6$, $18$, and $30$; toroids of $n=1$, $2$, $4$, and $6$; and ambient poloidal field strengths of $\alphab=1$ (left) and $\alphab=0$ (right). The spatial axes are labeled in units of $v_0t$,
with horizontal $\varpi$ axes exaggerated by factors as in Figure \ref{fig:7}.
}
\label{fig:8}
\end{figure*}

In this appendix, we show the density and column density profiles of the very magnetized bubbles in Figure \ref{fig:8}.

\section{Simplified hydrodynamic 1D model}
\label{sec:HD_analog}

A simplified model of this acceleration process can be constructed in 1D utilizing a hydrodynamical analog (useful for the radial equation of motion of the magnetized cold gas in the acceleration region), an adiabatic flow of $\gamma=2$, driven by a wind initially of velocity $v_0$ and sound speed $v_\mathrm{A0}=v_0/\MA$.
We further simplify the interaction with the very low-density ambient medium by considering it to be a vacuum (a more refined simplified model would consider it a low-density gas with an appropriate ambient $\gamma$ value).
In the absence of any characteristic problem length and time-scale, such hydrodynamic flows are self-similar.
Simplifying the low-density ambient as a vacuum requires looking for solutions configured purely by a rarefaction wave such as in, e.g.,\ Equations (4.77), (4.76), and (4.56) in \citet{torobook}. Beyond this expanding rarefaction wave, the original states of wind and ambient medium remain.
In our notations, the equation for the velocity can be summarized for the case of the relevant rarefaction wave as in \citet{torobook} Equation (4.56),
\begin{equation}
\label{eqn:toro_rarefaction_v}
v_r=\frac{2}{\gamma+1}\left(\frac{v_0}{\MA}+\frac{\gamma-1}{2}v_0+\frac{r}{t}\right)
=\frac{2}{3}\left(\frac{v_0}{\MA}+\frac{v_0}{2}+\frac{r}{t}\right)
\end{equation}
for $v_0(1-1/\MA)<r/t<v_0(1+[2/(\gamma-1)]/\MA)=v_0(1+2/\MA)$, with the flow unperturbed by this rarefaction wave beyond this acceleration region --- that is, free wind $v_r=v_0$ for $r/t<v_0(1-1/\MA)$ and vacuum for $r/t>v_0(1+2/\MA)$.
Here we could have written the velocity as simply $v$ and the space coordinate as $z$ instead of $r$ because in this 1D model curvature effects are not included.
Both the gradient value of $2/3t$ and the beginning and end of the acceleration region (both in position and velocity space) coincide with the analytical model presented in Section \ref{sec:method}. Moreover, Equation (\ref{eqn:toro_rarefaction_v}) can be applied to rarefaction waves in a more general context of interactions and boundary conditions to
the ambient medium and the free wind (changing its range of validity and perhaps also changing $v_0$ or $\MA$), and this shows that other 1D self-similar rarefaction waves would have the same gradient of acceleration, although the details of their beginning and end points in PV space would depend on details of their interaction with the ambient medium, and perhaps also with the pseudopulses. The estimate $\partial_r v_r=2/(3t)$ has therefore wider validity than the interaction with a vacuum, provided self-similarity is preserved.

\end{document}